\documentclass[article,longauth]{aa}
\pdfoutput=1
\usepackage[varg]{txfonts}
\bibpunct{(}{)}{;}{a}{}{,}
\usepackage{graphicx}
\usepackage{xspace}
\usepackage{hyperref}
\hypersetup{pdfauthor={Ranc et al.}, pdftitle={MOA-2007-BLG-197: Exploring the brown dwarf desert}}
\usepackage{epstopdf}
\usepackage{ulem}
\usepackage{color}
\def\ie{\textit{i.e.}\xspace}

\def\v{\,,}
\newcommand\Eq[1]{Eq.~(\ref{#1})}
\newcommand\Fig[1]{Fig.~\ref{#1}}
\newcommand\Tab[1]{Table~\ref{#1}}
\newcommand\Sec[1]{Sect.~\ref{#1}}

\newcommand\Msun{\mathrm{M}_\sun}
\newcommand\MJ{\mathrm{M_J}}

\newcommand\dix[1]{\times 10^{#1}}
\newcommand\MOA{MOA-2007-BLG-197\xspace}
\newcommand\MOAL{MOA-2007-BLG-197L\xspace}
\newcommand\MOALb{MOA-2007-BLG-197Lb\xspace}
\newcommand\ML{M}

\newcommand\MP{m_\mathrm{BD}}

\newcommand\DS{D_\mathrm{S}}
\newcommand\DL{D_\mathrm{L}}
\newcommand\FS{F_\mathrm{S}}
\newcommand\FB{F_\mathrm{B}}
\newcommand\vperp{v_\perp}

\newcommand\Ecin{E_{\mathrm{kin},\perp}}
\newcommand\Epot{E_{\mathrm{pot},\perp}}

\newcommand\RS{R_\mathrm{S}}

\newcommand\thE{\theta_\mathrm{E}}
\newcommand\thS{\theta_\mathrm{S}}
\newcommand{\tE}{t_\mathrm{E}}
\newcommand\pirel{\pi_\mathrm{rel}}
\newcommand\piS{\pi_\mathrm{S}}
\newcommand\piE{\pi_\mathrm{E}}
\newcommand\piEE{\pi_\mathrm{E,E}}
\newcommand\piEN{\pi_\mathrm{E,N}}

\newcommand\magJ{J}
\newcommand\magH{H}
\newcommand\magK{K}
\newcommand\magKs{K_s}

\def\tE{t_\mathrm{E}}
\def\uo{u_\mathrm{0}}
\def\to{t_\mathrm{0}}
\newcommand\mhjd[1]{$\mathrm{THJD}\simeq #1$}
\newcommand\mas{\mathrm{mas}}
\newcommand\au{\mathrm{AU}}

\newcommand\kpc[1]{\mathrm{kpc}^{#1}}

\newcommand\m[1]{\mathrm{m}^{#1}}
\newcommand\magn{\mathrm{mag}}

\begin{document}

\title{MOA-2007-BLG-197: Exploring the brown dwarf desert}

\author{C.~Ranc\inst{\ref{IAP},\ref{PLANET}}
\and A.~Cassan\inst{\ref{IAP},\ref{PLANET}}
\and M.~D.~Albrow\inst{\ref{Albrow},\ref{PLANET}}
\and D.~Kubas\inst{\ref{IAP},\ref{PLANET}}
\and I.~A.~Bond\inst{\ref{Bond},\ref{MOA}}
\and V.~Batista\inst{\ref{IAP},\ref{PLANET}}
\and J.-P.~Beaulieu\inst{\ref{IAP},\ref{PLANET}}
\and D.~P.~Bennett\inst{\ref{Bennett},\ref{MOA}}
\and M.~Dominik\inst{\ref{Dominik},\ref{PLANET}}
\and Subo~Dong\inst{\ref{Dong},\ref{mFUN}}
\and P.~Fouqu\'e\inst{\ref{Fouque},\ref{FouqueCFHT},\ref{PLANET}}
\and A.~Gould\inst{\ref{Gould},\ref{mFUN}}
\and J.~Greenhill$^\dagger$\inst{\ref{Greenhill},\ref{PLANET}}
\and U.~G.~J{\o}rgensen\inst{\ref{Jorgensen},\ref{PLANET}}
\and N.~Kains\inst{\ref{Sahu},\ref{Dominik},\ref{PLANET}}
\and J.~Menzies\inst{\ref{Menzies},\ref{PLANET}}
\and T.~Sumi\inst{\ref{Sumi},\ref{MOA}}
\and E.~Bachelet\inst{\ref{Bachelet},\ref{PLANET}}
\and C.~Coutures\inst{\ref{IAP},\ref{PLANET}}
\and S.~Dieters\inst{\ref{IAP},\ref{PLANET}}
\and D.~Dominis~Prester\inst{\ref{Dominis},\ref{PLANET}}
\and J.~Donatowicz\inst{\ref{Donatowicz},\ref{PLANET}}
\and B.~S.~Gaudi\inst{\ref{Gould},\ref{mFUN}}
\and C.~Han\inst{\ref{Han},\ref{mFUN}}
\and M.~Hundertmark\inst{\ref{Dominik},\ref{Jorgensen}}
\and K.~Horne\inst{\ref{Dominik},\ref{PLANET}}
\and S.~R.~Kane\inst{\ref{Kane},\ref{PLANET}}
\and C.-U.~Lee~\inst{\ref{Lee},\ref{mFUN}}
\and J.-B.~Marquette\inst{\ref{IAP},\ref{PLANET}}
\and B.-G.~Park~\inst{\ref{Lee},\ref{mFUN}}
\and K.~R.~Pollard\inst{\ref{Albrow},\ref{PLANET}}
\and K.~C.~Sahu\inst{\ref{Sahu},\ref{PLANET}}
\and R.~Street\inst{\ref{Street},\ref{PLANET}}
\and Y.~Tsapras\inst{\ref{Street},\ref{Wambsganss},\ref{PLANET}}
\and J.~Wambsganss\inst{\ref{Wambsganss},\ref{PLANET}}
\and A.~Williams\inst{\ref{WilliamsPerth},\ref{WilliamsBentley},\ref{PLANET}}
\and M.~Zub\inst{\ref{Wambsganss},\ref{PLANET}}
\and F.~Abe\inst{\ref{Abe},\ref{MOA}}
\and A.~Fukui\inst{\ref{Fukui},\ref{MOA}}
\and Y.~Itow\inst{\ref{Abe},\ref{MOA}}
\and K.~Masuda\inst{\ref{Abe},\ref{MOA}}
\and Y.~Matsubara\inst{\ref{Abe},\ref{MOA}}
\and Y.~Muraki\inst{\ref{Abe},\ref{MOA}}
\and K.~Ohnishi\inst{\ref{Ohnishi},\ref{MOA}}
\and N.~Rattenbury\inst{\ref{Rattenbury},\ref{MOA}}
\and To.~Saito\inst{\ref{Saito},\ref{MOA}}
\and D.~J.~Sullivan\inst{\ref{Sullivan},\ref{MOA}}
\and W.~L.~Sweatman\inst{\ref{Sweatman},\ref{MOA}}
\and P.~J.~Tristram\inst{\ref{Tristram},\ref{MOA}}
\and P.~C.~M.~Yock\inst{\ref{Yock},\ref{MOA}}
\and A.~Yonehara\inst{\ref{Yonehara},\ref{MOA}}
}

\institute{
\label{IAP} Sorbonne Universités, UPMC Univ Paris 6 et CNRS, UMR 7095, Institut d’Astrophysique de Paris, 98 bis bd Arago, 75014 Paris, France
\and \label{Albrow} University of Canterbury, Dept. of Physics and Astronomy, Private Bag 4800, 8020 Christchurch, New Zealand
\and \label{Bond} Institute of Natural and Mathematical Sciences, Massey University, Private Bag 102-904, North Shore Mail Centre, Auckland, New Zealand
\and \label{Bennett} Department of Physics, University of Notre Dame, Notre Dame, IN 46556, USA
\and \label{Dominik} SUPA, School of Physics \& Astronomy, North Haugh, University of St Andrews, KY16 9SS, Scotland, UK
\and \label{Dong} Kavli Institute for Astronomy and Astrophysics, Peking University, Yi He Yuan Road 5, Hai Dian District, Beijing 100871, China
\and \label{Fouque} IRAP, CNRS - Universit\'e de Toulouse, 14 av. E. Belin, F-31400 Toulouse, France
\and \label{FouqueCFHT} CFHT Corporation, 65-1238 Mamalahoa Hwy, Kamuela, Hawaii 96743, USA
\and \label{Gould} Department of Astronomy, Ohio State University, 140 W. 18th Ave., Columbus, OH  43210, USA
\and \label{Greenhill} School of Math and Physics, University of Tasmania, Private Bag 37, GPO Hobart, 7001 Tasmania, Australia
\and \label{Jorgensen} Niels Bohr Institutet, K{\o}benhavns Universitet, Juliane Maries Vej 30, 2100 K{\o}benhavn {\O}, Denmark
\and \label{Sahu} Space Telescope Science Institute, 3700 San Martin Drive, Baltimore, MD 21218, USA
\and \label{Menzies} South African Astronomical Observatory, PO Box 9, Observatory 7935, South Africa
\and \label{Sumi} Department of Earth and Space Science, Graduate School of Science, Osaka University, Toyonaka, Osaka 560-0043, Japan
\and \label{Bachelet} Qatar Environment and Energy Research Institute, Qatar Foundation, P.O. Box 5825, Doha, Qatar
\and \label{Dominis} Department of Physics, University of Rijeka, Radmile Matej v{c}i\'{c} 2, 51000 Rijeka, Croatia
\and \label{Donatowicz} Technical University of Vienna, Department of Computing,Wiedner Hauptstrasse 10, 1040 Wien, Austria
\and \label{Han} Department of Physics, Chungbuk National University, Cheongju 371-763, Korea
\and \label{Kane} Department of Physics and Astronomy, San Francisco State University, 1600 Holloway Avenue, San Francisco, CA 94132, USA
\and \label{Lee} Korea Astronomy and Space Science Institute, 776 Daedukdae-ro, Daejeon, Korea
\and \label{Street} Las Cumbres Observatory Global Telescope Network, 6740 Cortona Drive, suite 102, Goleta, CA 93117, USA
\and \label{Wambsganss} Astronomisches Rechen-Institut, Zentrum f{\"u}r Astronomie der Universit{\"a}t Heidelberg (ZAH), M\"onchhofstra{\ss}e 12-14, 69120 Heidelberg, Germany
\and \label{WilliamsPerth} Perth Observatory, Walnut Road, Bickley, Perth 6076, Australia
\and \label{WilliamsBentley} International Centre for Radio Astronomy Research, Curtin University, Bentley, WA 6102, Australia
\and \label{Abe} Solar-Terrestrial Environment Laboratory, Nagoya University, Nagoya 464-8601, Japan
\and \label{Fukui} Okayama Astrophysical Observatory, National Astronomical Observatory of Japan, 3037-5 Honjo, Kamogata, Asakuchi, Okayama 719-0232, Japan
\and \label{Ohnishi} Nagano National College of Technology, Nagano 381-8550, Japan
\and \label{Rattenbury} Department of Physics, University of Auckland, Private Bag 92019, Auckland, New Zealand
\and \label{Saito} Tokyo Metropolitan College of Aeronautics, Tokyo 116-8523, Japan
\and \label{Sullivan} School of Chemical and Physical Sciences, Victoria University, Wellington, New Zealand
\and \label{Sweatman} Institute of Information and Mathematical Sciences, Massey University at Albany, Private Bag 102904, North Shore 0745, Auckland, New Zealand
\and \label{Tristram} Mt. John University Observatory, P.O. Box 56, Lake Tekapo 8770, New Zealand
\and \label{Yock} Department of Physics, University of Auckland, Private Bag 92019, Auckland, New Zealand
\and \label{Yonehara} Department of Physics, Faculty of Science, Kyoto Sangyo University, 603-8555, Kyoto, Japan
\and \label{PLANET} PLANET/RoboNET Collaboration
\and \label{MOA} MOA Collaboration
\and \label{mFUN} $\mu$FUN Collaboration
}

\date{Received <date> / accepted <date>}

\abstract{
	We present the analysis of \MOALb, the first brown dwarf companion to a Sun-like star detected through gravitational microlensing. The event was alerted and followed-up photometrically by a network of telescopes from the PLANET, MOA, and $\mu\mathrm{FUN}$ collaborations, and observed at high angular resolution using the NaCo instrument at the VLT. From the modelling of the microlensing light curve, we derived basic parameters such as, the binary lens separation in Einstein radius units ($s \simeq 1.13$), the mass ratio $q=(4.732\pm0.020) \dix{-2}$ and the Einstein radius crossing time ($\tE \simeq 82\,\mathrm{d}$). Because of this long time scale, we took annual parallax and orbital motion of the lens in the models into account, as well as finite source effects that were clearly detected during the source caustic exit. To recover the lens system's physical parameters, we combined the resulting light curve best-fit parameters with $(J,H,K_s)$ magnitudes obtained with VLT NaCo and calibrated using IRSF and 2MASS data. From this analysis, we derived a lens total mass of $0.86 \pm 0.04\,\Msun$ and a lens distance of $\DL = 4.2 \pm 0.3\,\mathrm{kpc}$. We find that the companion of \MOAL is a brown dwarf of $41 \pm 2\,\MJ$ observed at a projected separation of $a_\perp = 4.3 \pm 0.1\,\au$, and orbits a $0.82 \pm 0.04\,\Msun$ G-K dwarf star. We then placed the companion of \MOAL in a mass-period diagram consisting of all brown dwarf companions detected so far through different techniques, including microlensing, transit, radial velocity, and direct imaging (most of these objects orbit solar-type stars). To study the statistical properties of this population, we performed a two-dimensional, non-parametric probability density distribution fit to the data, which draws a structured brown dwarf landscape. We confirm the existence of a region that is strongly depleted in objects at short periods and intermediate masses ($P\lesssim 30\,\mathrm{d}$, $M \sim 30-60\,\MJ$), but also find an accumulation of objects around $P \sim 500\,\mathrm{d}$ and $M \sim 20\,\MJ$, as well as another depletion region at long orbital periods ($P \gtrsim 500\,\mathrm{d}$) and high masses ($M \gtrsim 50\,\MJ$). While these data provide important clues on the different physical mechanisms of formation (or destruction) that shape the brown dwarf desert, more data are needed to establish their relative importance, in particular as a function of host star mass. Future microlensing surveys should soon provide more detections, in particular for red dwarf hosts, thus uniquely complementing the solar-type host sample.
}

\keywords{Gravitational lensing: micro - Planets and satellites: detection - Brown dwarfs}

\maketitle

\section{Introduction}

	Gravitational microlensing is a powerful technique for detecting extrasolar planets \citep{MaoPaczynski1991}, and it holds great promise for detecting populations of brown dwarf companions to stars. Compared to other detection techniques, microlensing provides unique information on the population of exoplanets, because it allows the detection of very low-mass planets (down to the mass of the Earth) at long orbital distances from their host stars (typically 0.5 to 10 AU). It is also the only technique that allows discovery of exoplanets and brown dwarfs at distances from the Earth greater than a few kiloparsecs, up to the Galactic bulge, which would have been hard to detect with other methods.
	
	Exoplanets are found to be frequent by all detection techniques \citep[e.g.,][]{Cassan2012,Bonfils2013,Mayor2011,Sumi2011,Gould4years2010}, and recent statistical microlensing studies even imply that there are, on average, one or more bound planets per Milky Way star \citep{Cassan2012}. Conversely, brown dwarfs appear to be intrinsically rare, to the point that shortly after the first exoplanet detections, it led to the idea of a ``brown dwarf desert'' \citep{Marcy2000} bridging the two well-defined regions of binary stars and planetary systems. While in the past, brown dwarfs were defined as objects of mass within the deuterium- and hydrogen-burning limits \citep[$13-74\,\MJ$,][]{Burrows2001}, it appears today that different formation scenarios can build objects with similar masses but with different natures (super-massive planets, or low-mass brown dwarfs). An object formed via core accretion and reaching $13\,\MJ$ would, for example, be able to start deuterium burning, as would an object of same mass formed by gravitational collapse of a cloud or in a protoplanetary disk \citep{MolMordasini2012}. 
	
	Despite their low occurrence, a number of brown dwarf companions to stars have been discovered by different methods: radial velocity and transit \citep[e.g.,][]{Moutou2013,Diaz2013,Sahlmann2011,Johnson2011,Deleuil2008}, direct imaging \citep[e.g.,][]{Lafreniere2007} and microlensing. With regards to microlensing, there are still a few brown dwarf detections, but this is mainly because until now observing priority has been given to exoplanets. Nevertheless, these detections are of particular interest because they provide a unique view of brown dwarfs around low-mass stars (mainly M dwarfs), which complements the currently available sample mostly composed of solar-type stars. 
	
	Microlensing detections so far include isolated brown dwarfs, brown dwarfs hosting planets, and brown dwarf companions to stars. The first isolated  brown dwarf  detected through  gravitational microlensing (a $59\pm4\,\MJ$ brown dwarf located at $525\pm40$ pc in the thick disk of the Milky Way) was reported by \cite{Gould2009} in microlensing event OGLE-2007-BLG-224.
Two brown dwarfs with planetary-mass companions were discovered in events OGLE-2009-BLG-151/MOA-2009-BLG-232 and OGLE-2011-BLG-0420 by \cite{Choi2013}. In both cases, the planets were super Jupiters ($7.9\pm0.3\,\MJ$ and $9.9\pm0.5\,\MJ$, respectively) with the hosts being low-mass brown dwarfs ($19\pm1\,\MJ$ and $26\pm1\,\MJ$, respectively),
with very tight orbits (below $0.4~\au$). Similarly, \cite{Han2013} report a $23\pm2~\MJ$ field brown dwarf hosting a $1.9\pm0.2~\MJ$ planet in a tight system after the analysis of the event OGLE-2012-BLG-0358.

	The first published microlensing detection of a brown dwarf companion to a star is OGLE-2008-BLG-510/MOA-2008-BLG-369, which was first reported by \cite{Bozza2012} as an ambiguous case between a binary-lens and a binary-source event. The data were reanalysed by \cite{Shin2012a}, who concluded that the binary-lens model involving a massive brown dwarf orbiting an M dwarf was preferred. 
\cite{Shin2012b} conducted a database search for brown dwarf companions by focusing on microlensing events that exhibit low mass ratios. Among seven good candidates with well-determined masses (combination of Einstein radius and parallax measurements), they found two events that involve brown dwarfs: OGLE-2011-BLG-0172/MOA-2011-BLG-104 with mass $21\pm10\,\MJ$ around an M dwarf, and MOA-2011-BLG-149, a $20\pm2\,\MJ$
brown dwarf also orbiting an M dwarf. Similarly, \cite{Bachelet2012b} reported the detection of another $\sim 52\,\MJ$ brown dwarf orbiting an M dwarf in MOA-2009-BLG-411, although the lens mass could not be determined exactly, and was estimated through statistical realisations of Galactic models. In microlensing event MOA-2010-BLG-073, \cite{Street2013} find that the lens was composed of a $11.0\pm2.0~\MJ$ companion (hence near the planet/brown boundary) orbiting an M dwarf of $0.16\pm0.03~\Msun$. \cite{Jung2014} report the detection of a star at the limit of the brown dwarf regime hosting a companion at the planet/brown dwarf boundary ($13\pm2\,\MJ$). More recently, \cite{Park2015arXiv} have reported the discovery of a binary system composed of a $33.5 \pm 4.2 \, \MJ$ brown dwarf orbiting a late-type M dwarf in microlensing event OGLE-2013-BLG-0578.

	Here we report the first microlensing discovery of a brown dwarf orbiting a Sun-like star. This new brown dwarf has a mass of $42\,\MJ$ and it was observed at a projected separation of $4.3\,\au$ from its G-K dwarf host star. 
	
	In sec. 2, we present photometric data collected on \MOAL by several round-the-world telescopes, as well as high resolution adaptative optics images taken with NaCo at VLT. In sec. 3, we present the full analysis of the light curve and in sec. 4 we derive the physical parameters of the lens by combining all independent informations. In sec. 5, we adopt a statistical point of view to analyse the current population of brown dwarfs detected with different methods using non-parametric probability density estimation tools. In sec. 6 we summarise our results, and underline the importance of future microlensing observations to characterise the populations of objects in the mass region between planets and stars.

\section{Observational data}

\subsection{Alert and follow-up} \label{par:followup}

\begin{figure*}[t]
\begin{center}
\includegraphics[scale=1]{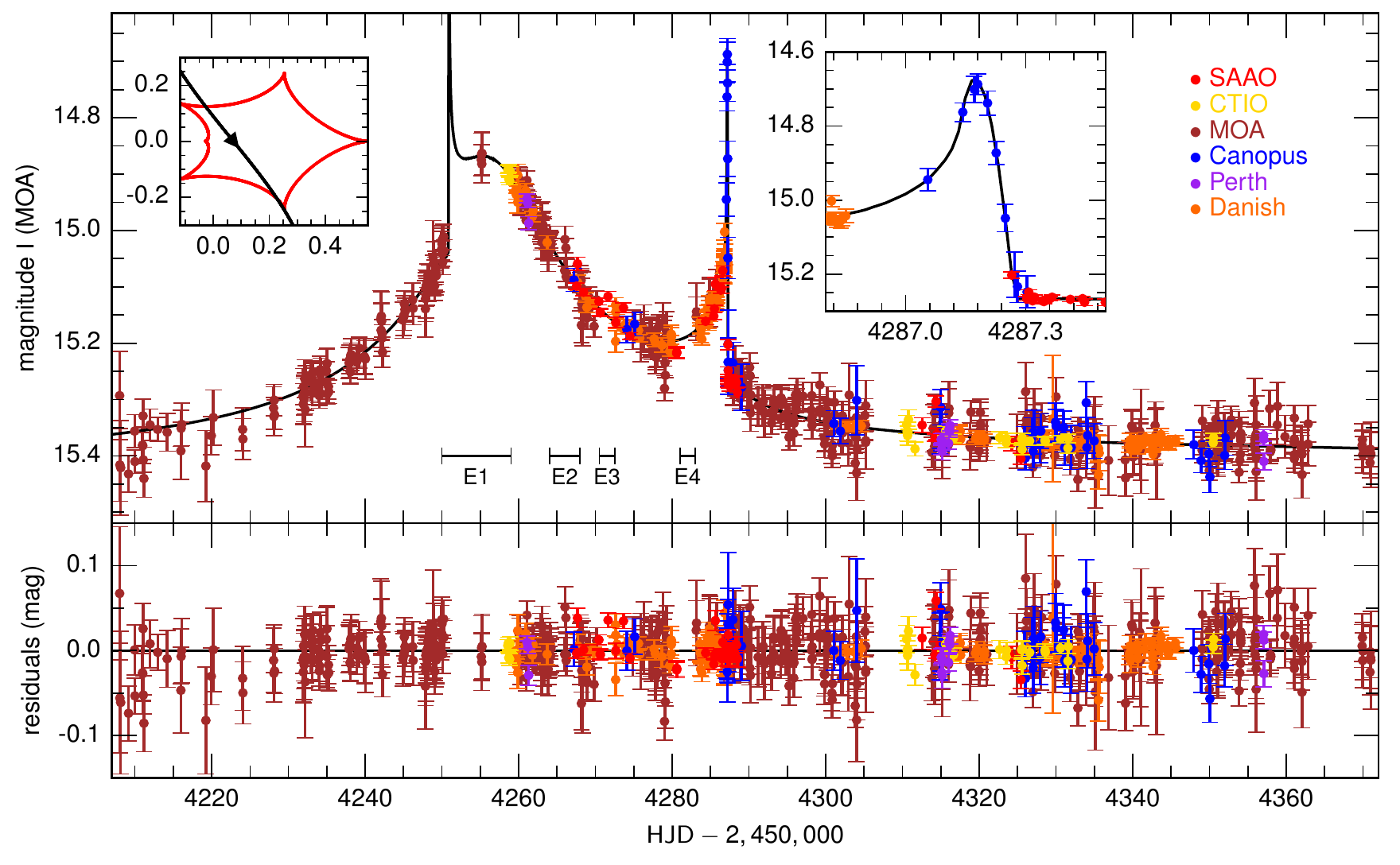}
\caption{In the upper panel, the light curve of \MOA and the best-fit model (solid line)  are plotted with a zoom on the caustic exit on the right-hand side. On the left-hand side, the structure of the resonant caustic is drawn in red, as well as the trajectory of the source, in black (axes are in Einstein radius units). The source is too small to be distinguished. The four intervals $E_{1-4}$ indicate time intervals where possible source caustic crossing (caustic entry) have been investigated. In the lower panel, the residuals of the best-fit model are shown.}
\label{fig:lc}
\end{center}
\end{figure*}

	\MOA ($l=359.711$, $b=-5.509$, or RA (J2000) = 18:07:04.729, DEC (J2000) = -31:56:46.77) is a microlensing event that was alerted by the MOA collaboration (1.8m telescope located in Mount John, New Zealand) in 2007 May 28 (or \mhjd{4249}\footnote{$\mathrm{THJD}=\mathrm{HJD}-2,450,000$.} on the photometric light curve shown \Fig{fig:lc}). Soon after, between 1-5 June (\mhjd{4251}-4255), a very cloudy weather in New Zealand seriously affected the quality and reliability of the photometry, resulting in gaps in the time series. A magnification peak in the MOA light curve was passed around June 6 (\mhjd{4258}). On June 8 (\mhjd{4259.7}), the PLANET collaboration added the event to its target list  as a regular mid-magnification object. Follow-up observations include data from the PLANET network using the Danish telescope 1.54m at La Silla (Chile), Canopus 1m in Hobart (Tasmania),  SAAO 1m in Sutherland (South Africa) and Perth 0.6m telescope (Australia), while the $\mu$FUN collaboration collected data from CTIO 1.3m in Mount Cerro Tololo (Chile). 
   
   On June 19  (\mhjd{4272}), PLANET observers noticed that a few MOA data points were slightly above the standard single-lens theoretical curve. As this usually happens quite often in microlensing data, no alert was released. Indeed SAAO soon revealed that the alleged deviation was a false alert, but follow-up observations were continued. On July 5 (\mhjd{4287}), even though the full moon affected the quality of the observations, a public alert was issued after Danish 1.54m data were found to be above the single lens curve by more than 0.2 mag for more than five consecutive days. Intensive follow-up observations from SAAO confirmed the rise in brightness, announcing a caustic crossing. While Perth was overclouded, Canopus then took over, and was the only telescope to densely cover the caustic exit, which took place during the night of July 4 in Australia (\mhjd{4287.0}-4287.4, see inset of \Fig{fig:lc}).
   
\begin{table*}[t]
\centering
\caption{Telescopes and photometric data sets.}
\begin{tabular}{lllll}
\hline\hline
Telescope & Location & Filter & Data\tablefootmark{a}  & $f$\tablefootmark{b} \\
\hline
MOA	 ($1.8\m{}$)  & Mount John, New Zealand  & $R_M$\tablefootmark{c} & 504 & 1.4\\
PLANET Danish ($1.54~\mbox{m}$)  & La Silla, Chile & $I$ & 167  & 1.7 \\
PLANET Canopus ($1.0~\mbox{m}$) & Mount Canopus, Tasmania & $I$ & 45 & 3.0 \\
PLANET SAAO ($1.0~\mbox{m}$) & Sutherland, South Africa & $I$  & 43 & 1.6 \\
$\mu\mathrm{FUN}$ CTIO ($1.3~\mbox{m}$) & Mount Cerro Tololo, Chile & $I$ & 28 & 1.3 \\
PLANET Perth ($0.6~\mbox{m}$)  & Perth, Australia & $I$ & 15 & 1.4 \\
\hline
\end{tabular}
\tablefoot{\tablefoottext{a}{Number after data cleaning.}\tablefoottext{b}{Error bar rescaling factor (\Sec{par:photo-lc}).}\tablefoottext{c}{MOA broad R/I filter.}}
\label{tab:observations}
\end{table*}

\subsection{Photometric light curve} \label{par:photo-lc}

   The photometric data was reduced several times using different software to check their consistency. The final lightcurve data from all telescopes were extracted with the PLANET pipeline PySIS \citep{Albrow2009}, which is a DIA-based algorithm \citep[Difference Image Analysis, ][]{ISIS,AlardLupton}. For images taken in particularly bad weather conditions, we examined by eye each image in order to check whether the subtraction was correct or not. In the latter case we had to exclude them, but on the basis of image quality only. In the post-processing of the reduced data, we applied a cut in seeing and sky background, although  in a very conservative way so that possible low-amplitude signals were not rejected. Not surprisingly, most of the data taken by MOA during the very bad weather period mentioned before (\mhjd{4251}-4255) had to be discarded. Unfortunately, a critical light curve feature is thereby not covered (which has important consequences, see \Sec{sec:lc-models}), however there was no other choice but to remove these data, as they might otherwise have affected the reliability of our models.

   The final light curve data amounts to a total of 802 data points. They are summerised in \Tab{tab:observations}. As seen in the table, all telescopes use a similar $I$-band filter, apart from the MOA 1.8m telescope which is equipped with a broad R/I filter (referred to as $R_M$). Additionally, a few $V$-band images were taken by PLANET and $\mu$FUN to produce colour-magnitude diagrams (see \Sec{sec:physical-parameters-source}). 

   The last concern about the photometry was the estimation of the error bars of the data. Galactic bulge fields are highly crowded with stars, and during a microlensing event, the flux variation can easily span two order of magnitudes for high-magnification events. These pose severe challenges to manage a good estimation of the error bars. As a matter of fact, in microlensing experiments it is long known that data reduction software usually underestimates error bars. Furthermore, error bars can vary significantly from one data set to another, with the risk that one data set dominates over the others at the modelling stage. A relatively robust method to prevent these drawbacks is to rescale the error bars, based on the best model fitting the data. For each data set, the (classical) $\chi^2$ is set up to the number of degrees of freedom by adjusting a rescaling factor $f$ in the formula $\sigma'^2=f^2 \sigma^2 + \sigma_0^2$, where $\sigma'$ and $\sigma$ are respectively the rescaled and initial error bars on the magnitudes, and $\sigma_0=4\dix{-4}$ a constant accounting for the data most highly magnified. The $f$ factors are given in \Tab{tab:observations} for each data set.

\subsection{VLT NaCo high resolution images} \label{sec:datanaco}

\begin{figure}[ht]
\begin{center}
\includegraphics[scale=1]{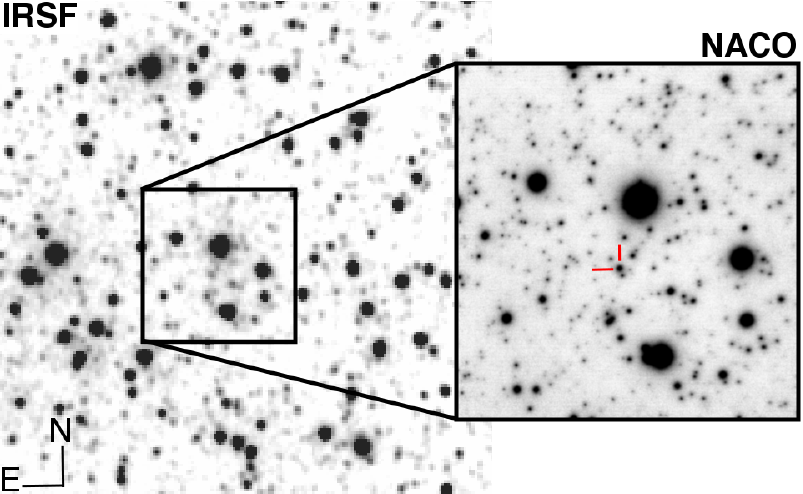}
\caption{The image on the left shows a sub-region ($94\arcsec \times 77\arcsec$) of an original $3\arcmin\times3\arcmin$ $\magKs$-band IRSF image used to calibrate the NaCo magnitudes. The image on the right is the corresponding $\magKs$-band NaCo image ($27\arcsec \times 25\arcsec$) used to cross-identify the stars.}
\label{fig:nacoirsf}
\end{center}
\end{figure}

	On the night 20/21 of August 2007 (\mhjd{4333.0}) we obtained first epoch observations\footnote{ESO Programme ID 279.C-5044(A).} of high resolution adaptive optics (AO) images in the near-infrared bands $\magJ$, $\magH$ and $\magKs$ using the NaCo instrument, mounted on 8.2m ESO VLT Yepun telescope (\Fig{fig:nacoirsf}). The source star was then still magnified by a factor of about $A \simeq 1.30$. A year later, on the night 2008 August 3/4  (\mhjd{4682.1}), we carried out second epoch observations\footnote{ESO Programme ID 381.C-0425(A).} while the event was back to baseline magnitude. In principle, when two epochs are obtained at different magnifications, it should be possible to directly disentangle the lens flux from the source flux. In our case, however, the combination of a relatively high blending factor and low magnification did not support this direct measurement. Therefore,  in the global analysis we used only the first epoch images. We reduced and calibrated the NaCo images following the general method outlined in \cite{Kubas2012} and briefly described below.
		
The night was clear and stable according to the observatory night logs, and the target was observed at low airmass and in good seeing conditions ($< 0.8\arcsec$). The data were taken in auto-jitter mode within a $10\arcsec$ jitter box including the target in order to be able to correct for bad pixels and sky background. The AO correction was done using  a $\magK = 11.17 \pm 0.02\ \magn$ star (2MASS 18070464-3156423) at an angular separation of about $5\arcsec$ from the microlensing target.  The data were dark subtracted, flat-fielded and co-added with the tools underlying the NaCo pipeline software \citep{Devillard1999}.

   To derive the photometry from the reduced data, the first step was to compute the zeropoints  for the conversion of instrumental magnitudes to calibrated magnitudes. A first possibility was to use catalogued stars within the target frame field-of-view  (FOV), in our case $28\arcsec \times 28\arcsec$. While several cross-matches between the NaCo frames and the 2MASS catalogue were identified, only 2MASS 18070520-3156409 ($J=14.24 \pm 0.05$, $H=13.69  \pm 0.05$, $K=13.48  \pm 0.05$) turned out as suitable (other potential calibrators in the NaCo images were either saturated or in the nonlinear regime of the detector). We finally checked that this star was not variable, by comparing a series of $H$-band images taken with Andicam at CTIO, which is well calibrated to 2MASS thanks to its $2.4\arcmin \times 2.4\arcmin$ FOV. A second option for calibration was to use the zeropoints derived from the photometric standards taken with NaCo directly before the observations on the night  20/21 of August and at similar airmass. For calibration we used $\magJ, \magH, \magK$ magnitudes\footnote{Note that the difference in the transmission profile between K and Ks bands is less than $1 \%$, so negligible in the present case.} of star 9160-S870-T in the listed standards of \cite{Persson1998}, which had the following advantages: this star was brighter than the previous 2MASS reference, and the smaller pixel scale made it less sensitive to blending contamination in our crowded field. For consistency, however, we measured the magnitude of 2MASS 18070520-3156409 in the NaCo frame, and found an agreement to better than $3\%$. 
   
	The second step was to extract accurate photometry from AO images in the infrared. This was not a straightforward task, since the shape of the point-spread function (PSF) often is not well fitted by analytical profiles, and also depends on the position of the target with respect to the star used for AO correction. 
Following \cite{Kubas2012}, we constructed a PSF reference directly from stars in the NaCo frame, using the StarFinder package \citep{2000SPIE.4007..879D}. This software was especially designed for AO images of crowded stellar fields. We found $\magJ=17.68\pm0.06$, $\magH=17.05 \pm 0.05$ and $\magKs=16.87 \pm 0.05$. The quoted error bars are dominated by the uncertainties in determining the true PSF shape and the scatter of the sky and unresolved background sources. The measurements obtained with StarFinder are summarised in \Tab{tab:NaComags}.
	
\begin{table}[!htbp]
\begin{center}
\begin{tabular}{lcccc}
\hline
\hline
Band & Epoch & Magnitude & Date [THJD] & FWHM \\
\hline
$\magJ$ & 1 & $17.68 \pm 0.06$ & 4333.03906250 & 0.19'' \\
$\magJ$ & 2 & $17.67 \pm 0.05$ & 4682.12500000 & 0.17'' \\
\hline
$\magH$ & 1 & $17.05 \pm 0.05$ & 4333.05468750 & 0.14'' \\
$\magH$ & 2 & $17.04 \pm 0.04$ & 4682.14453125 & 0.11'' \\
\hline
$\magKs$ & 1 & $16.87 \pm 0.05$ & 4333.02343750 & 0.11'' \\
$\magKs$ & 2 & $16.89 \pm 0.04$ & 4682.10546875 & 0.11'' \\
\hline
\end{tabular}
\end{center}
\caption{Derived NaCo magnitudes of the microlensing event target in the ($\magJ, \magH,\magKs$) filters for the two epochs 2007 August 20/21 (epoch 1) and 2008 August 3/4 (epoch 2). The values of Epoch 1 are used to constrain the lens mass-distance relation.}
\label{tab:NaComags}
\end{table}

\begin{figure}[t]
\centering
\includegraphics[width=8.5cm]{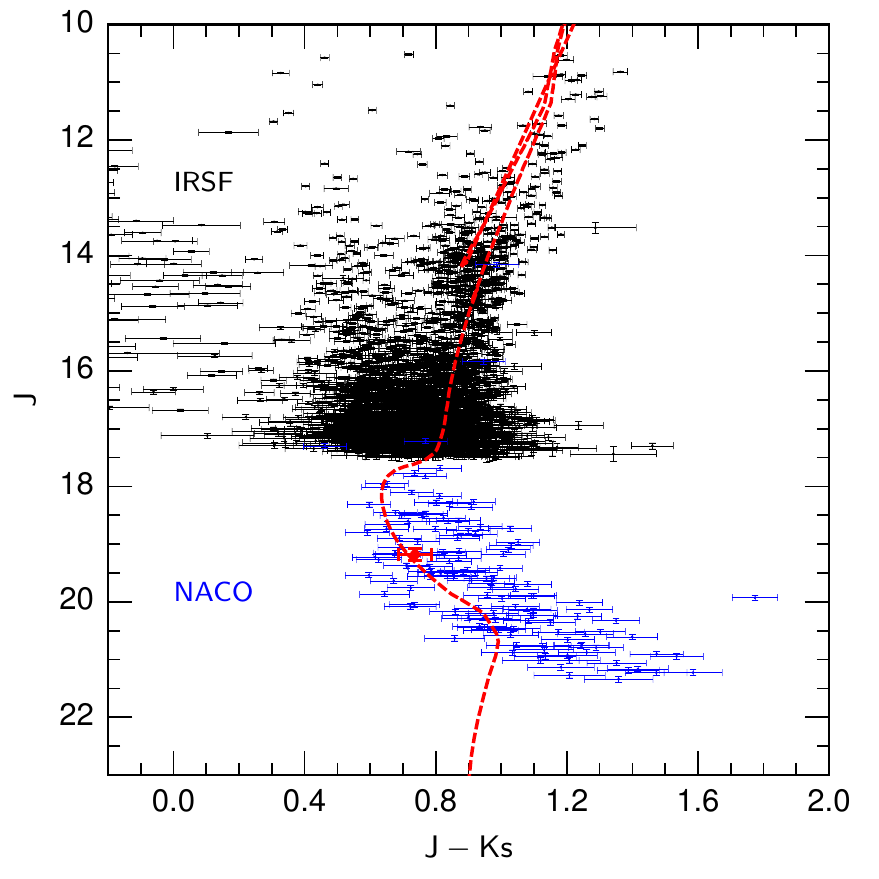}
\caption{$(\magJ,\magJ-\magKs)$ colour-magnitude diagram in the 2MASS system combining NaCo and IRSF data (respectively, in blue and black). The red dot is the magnitude/colour of the source derived from the source characterisation (\Sec{sec:physical-parameters-source}). The values are not corrected for interstellar extinction. The red clump (RCG) is fitted by the red over-plotted $10~\mbox{Gyr}$, solar-metallicity isochrone from \cite{Bressan2012}.}
\label{fig:cmdNaCo}
\end{figure}

	The final step consisted in correcting the target for interstellar extinction, by fitting the position of the red  clump giants (RCG) in the three colour-magnitude diagrams (CMD) involving the measured $\magJ, \magH,\magKs$ reddened magnitudes. Because the nominal range in magnitudes from NaCo is above the 2MASS faint limit, we performed InfraRed Survey Facility (IRSF) observations at SAAO to extend the available 2MASS star list into the regime of stars measured within the NaCo frame. We used \cite{Kato2007} to obtain the calibration of IRSF images with respect to the 2MASS reference-star catalogue, noting that \cite{Janczak2010} found that no additional colour term is needed between NaCo and IRSF filters \citep[the full process to build the calibration ladder is detailed in][]{Kubas2012}. Data to construct the NaCo+IRSF CMD are extracted from 2304 stars identified within a $3\arcmin$ circle around the target in IRSF images, and 135 stars identified in the NaCo images (\Fig{fig:nacoirsf}). The resulting de-reddened and calibrated CMD is plotted in \Fig{fig:cmdNaCo}, with a fit of the Red Clump Giant (RCG) position using a $10~\mbox{Gyr}$, $Z=0.019$ isochrone from \cite{Bressan2012}, and assuming a RCG distance modulus of $\mu=14.6\,\mathrm{mag}$. This distance modulus is longer than the value used in \cite{Kubas2012} since the event is located in the Galactic bar (see \Sec{sec:physical-parameters-source} for a detailed discussion). From this analysis, we derived the following extinctions: $A_\magJ=0.51\pm0.05\,\mathrm{mag}$, $A_\magH=0.33\pm0.05\,\mathrm{mag}$ and $A_{\magKs}=0.22\pm0.05\,\mathrm{mag}$. These data are used in \Sec{sec:NACOmassdist} to constrain the lens and mass distance.

\section{Light curve modelling} \label{sec:lc-models}
	
\subsection{Binary lens parameters} \label{sec:lc-models-ESBL}

	We start by modelling the light curve of \MOA with a static binary-lens model, in order to identify broad classes of possible solutions. In this model, the lens is characterised by its binary mass ratio $q$, and $s$, the projected separation of the binary lens in angular Einstein radius units $\thE$. Here,
\begin{equation}
   \thE = \sqrt{\frac{4GM}{c^2\DS}\left( \frac{\DS}{\DL} -1 \right)}\v
   \label{eq:thE}
\end{equation}
where $G$ is the gravitational constant, $c$ the speed of light, $\DL$ and $\DS$ are respectively the observer-lens and observer-source distances and $M$ the total mass of the lens. Four additional parameters describe the source mean rectilinear motion: the minimum impact distance $\uo$ between the source and the origin (here, the centre of mass of the system, with the more massive body on the right-hand side), $\to$, the date at which the source reaches $\uo$, $\tE$, the time it takes for the source to travel one Einstein angular radius, and $\alpha$, the angle between the source trajectory and the lens symmetry axis. 

	When the source approaches a caustic, finite-source effects cannot be neglected and substantial deviations from a point-source model are expected. Finite-source effects are included in the modelling through
\begin{equation}
   \rho = \frac{\thS}{\thE}\v
   \label{eq:rho}
\end{equation}
where $\thS$ is the angular radius of the source, and $\rho$ the same quantity but in Einstein radius units. Here and in the following, a point-source model is used when the source is far enough from the caustics ($t\leq 4240.0$ and $t\geq 4300.0$); closer to the caustics, a hexadecapole approximation to the finite-source magnification \citep{Gould2008} is used  ($t \in [4240.0, 4247.9] \cup [4260.0, 4286.0] \cup [4287.5, 4300.0]$), until it breaks down, approximately when the source is closer than $\sim 3\rho$ from the caustics. In this case, a full integration along the source images contours \citep{Dominik2007cont,Bozza2010,GouldGaucherel1997} or ray-shooting of the images are required \citep{Dong2006,Dong2009,Bennett2010} to compute the magnification. Since finite-source effects dramatically increase the computational cost, these time intervals are reduced as much as possible (here, $t \in [4247.9, 4260.0] \cup [4286.0, 4287.5]$).

	Microlensing light curves are usually sensitive to stellar limb darkening \citep[e.g.,][]{Albrow1999LD,Cassan254}. We thus model the source as a linear limb-darkened disk \citep{An2002,Zub2011} described by the intensity-normalised profile
\begin{equation}
   I(r) = \frac{1}{\pi}\left[ 1- \Gamma \left( 1-\frac{3}{2} \sqrt{1-r^2} \right) \right]\v
\end{equation}
where $0\leq r\leq 1$ is the fractional radius, and $\Gamma$ the linear limb-darkening (LLD) coefficient. While in special cases limb-darkening laws beyond the linear law may slightly improve the model \citep[e.g.,][]{Cassan69letter,Kubas2005}, in the case of \MOAL (relatively sparse data coverage of the caustic crossings) it is an excellent approximation. As seen in the right inset of \Fig{fig:lc}, only Canopus data are sensitive to limb darkening, although the light curve sampling is not dense enough to provide strong constraints on the LLD coefficient. We therefore use \cite{Claret2011} for the source surface gravity and effective temperature found in \Sec{sec:physical-parameters-source} ($\log{g}\sim4.47$, $T_\mathrm{eff}\sim 5350~\mbox{K}$) to estimate \MOAL LLD coefficients, and found $\sim0.4$ for the $I$ filter and $\sim0.5$ for the $R$ filter. We checked that refining the Canopus-$I$ LLD coefficient with a fit to the data leads to $\Gamma_I=0.48$. We then adopted $\Gamma_I=0.48$ for all $I$-band data and $\Gamma_{R_M}= 0.5$ (MOA $\Gamma_{R_M}$ filter is a broad R/I filter).

	Finally, two additional parameters, $\FS^i$ (source flux) and $\FB^i$ (blended flux), are included per individual observatory or filter, so that for a given individual data set $i$ the total flux of the microlensing target reads
   \begin{equation}
   F^i(t) = A(t)\, \FS^i + \FB^i\v
\end{equation}
where $A(t)$ is the time-dependent source flux magnification factor. The blending flux $\FB$ accounts for all luminous contributions other than the source, in particular, it includes the flux from the lens.
For the best model presented in \Tab{tab:lc-parameters}, the blending flux ratios $\FB/\FS$ are respectively $17.79$ (MOA), $38.14$ (SAAO), $12.85$ (CTIO), $3.97$ (Canopus), $66.97$ (Perth), $12.10$ (Danish). The high value for Perth is easily explained by the fact that the aperture of the telescope is small and the seeing was relatively high, which results in a higher blend fraction.

\subsection{Caustic-crossing parametrisation} \label{sec:causfix}
 
	The shape of the \MOA light curve (\Fig{fig:lc}) clearly indicates that the source crosses (\ie, exits) a fold caustic at \mhjd{4287.2}, mainly thanks to Canopus data densely covering this part of the light curve. On the other hand, the date of the caustic entry is ambiguous. In fact, for such a small angular source size as suggested by the caustic exit, the caustic entry could easily fit in one of the several gaps in the data coverage, in particular those marked as $E_{1-4}$ in \Fig{fig:lc}. 

	One of the modelling challenges for this event was to find  all possible local minima involving all possible caustic entry dates. The problem with the set of parameters $(\uo, \alpha, \tE, \to)$ described in the previous section is that it is not well suited to explore efficiently the parameter space. In fact, only very special combinations of these parameters produce caustic crossings at the right locations on the light curve. \cite{Albrow1999} proposed for the first time to introduce specific parameters to model a caustic crossing. \cite{Cassan2008} developed further this approach, and generalised it to a pair of caustic crossings (entry and exit) by introducing a specific parametrisation of the caustic curves. This method is particularly efficient for an event like \MOA. 
	
	In this formalism, two alternative parameters $t_{\rm in}$ and $t_{\rm out}$ are used to fit for the dates of the source entry (in) and exit (out), while two other parameters $s_{\rm in}$ and $s_{\rm out}$ (which are curvilinear distances along the caustic curve) are used to fit for the source centre ingress and egress points on the caustic. It has been demonstrated that this parametrisation is more efficient in locating all possible fitting source-lens trajectories \citep{Kains2009,Kains2012}, because they all produce caustic-crossing features at the observed dates. \cite{Cassan2010} later derived Bayesian priors on parameters $(t_{\rm in}, t_{\rm out}, s_{\rm in}, s_{\rm out})$ to explore even more efficiently the parameter space. The best-fit values derived from the minimisation process are finally converted back to the classical parameters. 
	
\begin{figure*}[ht]
\begin{center}
\includegraphics[width=\textwidth]{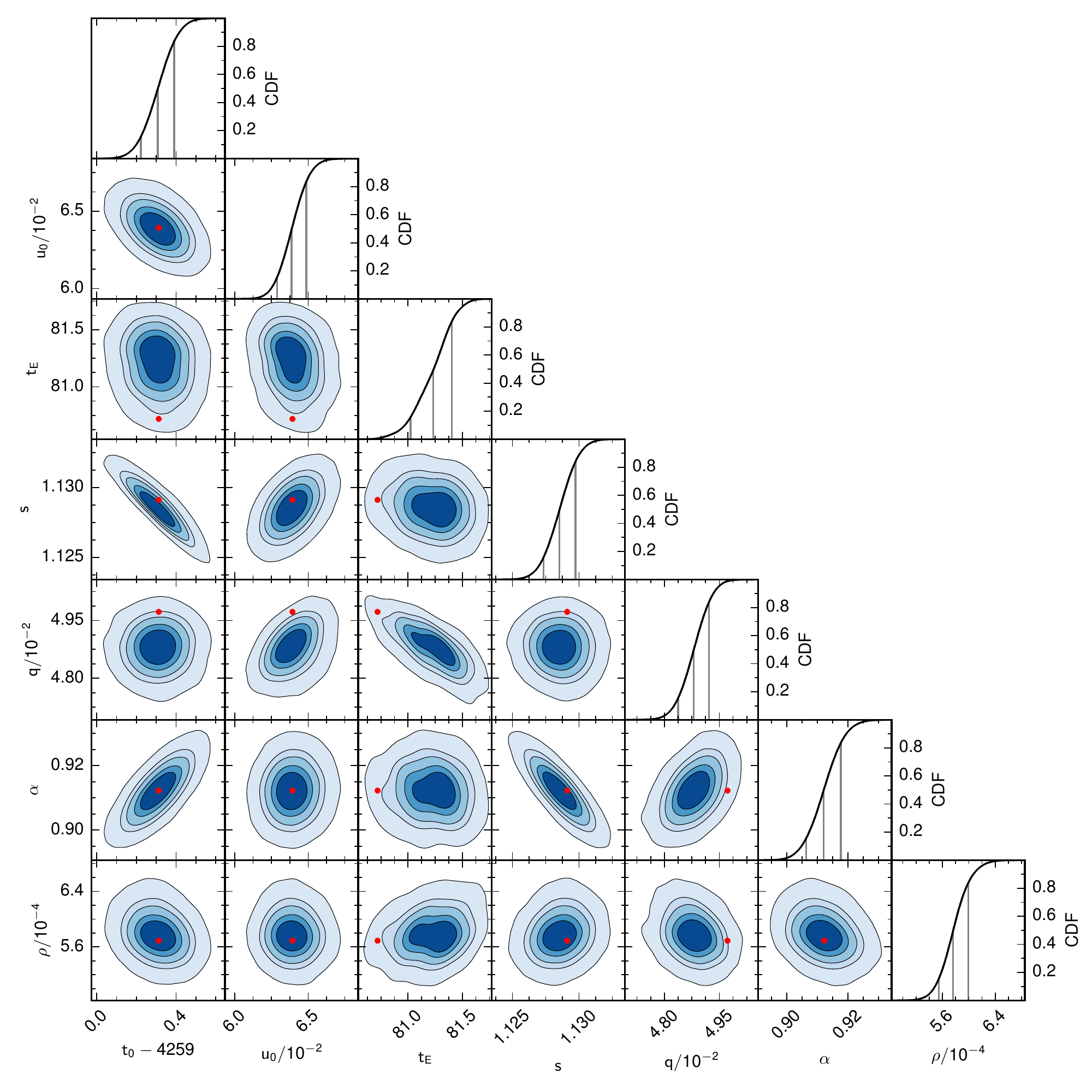}
\caption{Correlations between the parameters derived from ESBL's model discussed in \Sec{sec:lc-models-ESBL}. The red point refers to the best-fitting model obtained.}
\label{fig:ESBL}
\end{center}
\end{figure*}

	We explored\footnote{These very intensive computations were performed on the Tasmanian cluster TPAC (Tasmanian Partnership for Advanced Computing).} all static binary-lens models using this parametrisation for all possible source caustic crossing dates within the $E_1$ to $E_4$ intervals displayed in \Fig{fig:lc}, and for a regular $(\log s, \log q)$ grid of $30\times 30$ spanning $\log(0.4)\leq \log s \leq \log(4)$ and $-4\leq \log q \leq 0$. The minimisation was performed using a classical Markov Chain Monte Carlo (MCMC) algorithm. The best model involves a caustic entry in the time interval $E_1$, the longest of the four gaps explored. The best-fit values of the model parameters and the corresponding $\chi^2$ are listed in column ESBL (Extended-Source, Binary-Lens) of \Tab{tab:lc-parameters}, and posterior probability densities (correlation plots) are presented in \Fig{fig:ESBL}.

\subsection{Annual parallax} \label{sec:lc-models-ESBL-P}

	Given the long timescale found previously for the static binary lens model ($\tE\sim80~\mbox{days}$), this event is \textit{a priori} likely to exhibit annual parallax effects. The relative lens-source (annual) parallax is expressed as
\begin{equation}
	\pirel   = \frac{\mbox{AU}}{\DL}  - \frac{\mbox{AU}}{\DS}\v
\label{eq:pirel}
\end{equation}
so that the Einstein angular radius \Eq{eq:thE} also reads 
\begin{equation}
	\thE=\sqrt{\kappa \ML \pirel}\v
\label{eq:thEpirel}
\end{equation}
where $\kappa \simeq 8.144~\mbox{$\mas/{\rm M}_\sun$}$. The Einstein parallax vector $\boldsymbol{\pi_E}$ has an amplitude
\begin{equation}
	\piE = \frac{\pirel}{\thE}\v
\label{eq:piE}
\end{equation}
and a direction along the lens-source proper motion. As fitting parameters, it is convenient to decompose $\boldsymbol{\pi_E}$ into a northwards component $\piEN$ and an eastwards component $\piEE$ \citep{An2002}, with advantages discussed in \cite{Gould2004}. 

\begin{figure*}[ht]
\begin{center}
\includegraphics[width=\textwidth]{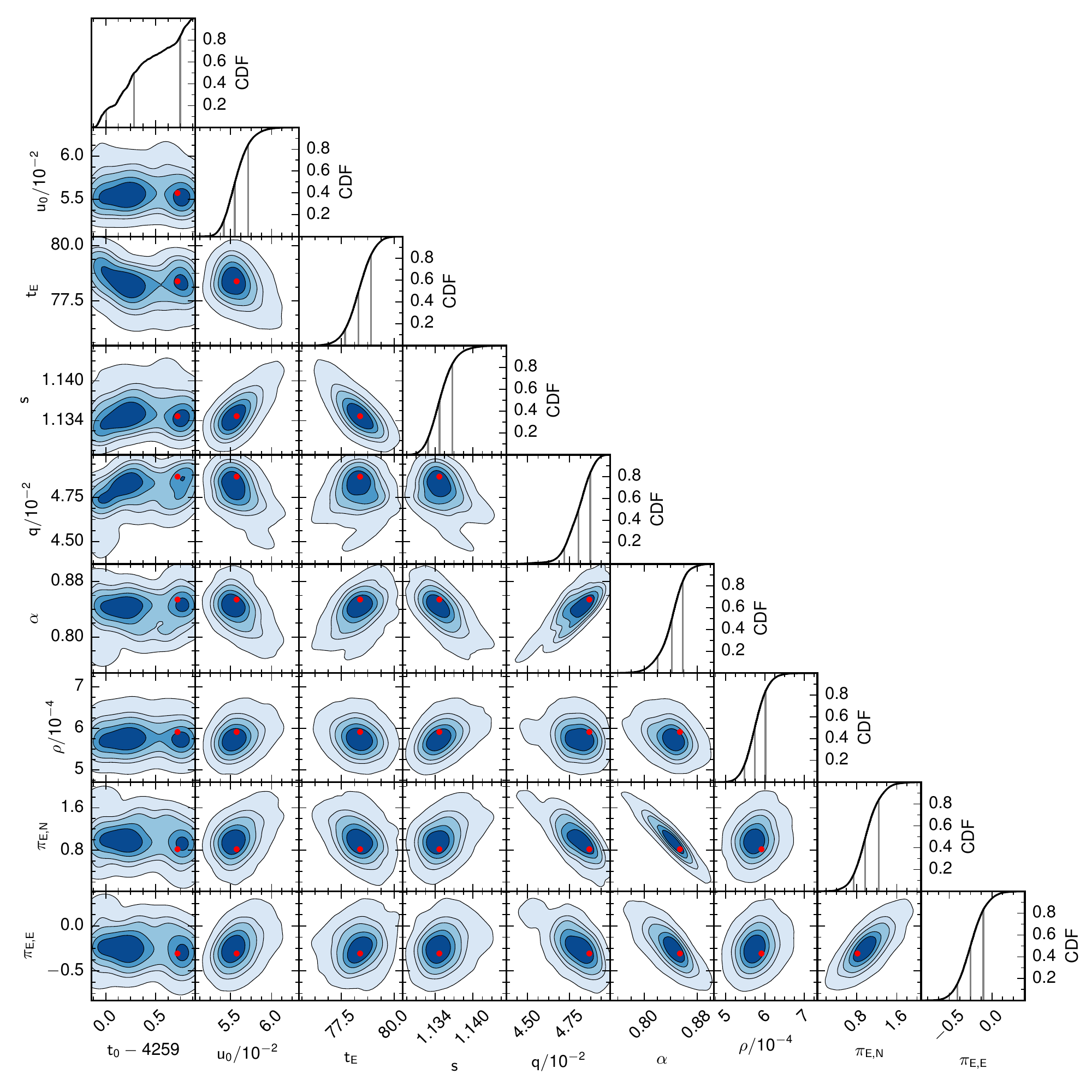}
\caption{Idem for ESBL+P's model.}
\label{fig:ESBLP}
\end{center}
\end{figure*}

	The best-fit model including parallax only (besides the parameters described in \Sec{sec:lc-models-ESBL}) is presented in column ESBL+P of \Tab{tab:lc-parameters}, and correlation plots are shown in \Fig{fig:ESBLP}. It can be noticed that the binary mass ratio $q$ and separation $s$ change little by including parallax compared to ESBL, resulting in a very similar resonant caustic structure. As expected, including parallax in the model improves the $\chi^2$, by around 100. Nevertheless, we found that several solutions with very different values of parallax (between $0.8$ to $1.5$) give almost identical $\chi^2$ differing by only one or two units. The differences between these models come from differences in the caustic entry date $t_{\rm in}$, \ie precisely where no data are available to constrain the model. Hence, although parallax improves the fit, it is unlikely that a workable measurement can be obtained (we discuss this further later).

\subsection{Orbital motion} \label{sec:lc-models-ESBL-OM}

	Orbital motion of the lens is also \textit{a priori} likely to produce noticeable effects, because the event is fairly long and the caustic resonant. The orbital rotation of the two lens components affects the caustic in two ways: it changes the projected binary separation $s$ and the orientation of the caustic in the plane of the sky. To first order, one can write $s(t) \simeq s + \dot s\,(t-t_r)$ and $\alpha(t) \simeq \alpha + \dot{\alpha}\,(t-t_r)$, where $t_r$ is an arbitrary reference date chosen close or equal to $t_0$. These two effects are included in the model through parameters $\gamma_\parallel = \dot{s}/s$ and $\gamma_\perp = \dot{\alpha}$, with $\gamma^2=\gamma_\parallel^2+\gamma_\perp^2$.

\begin{figure*}[ht]
\begin{center}
\includegraphics[width=\textwidth]{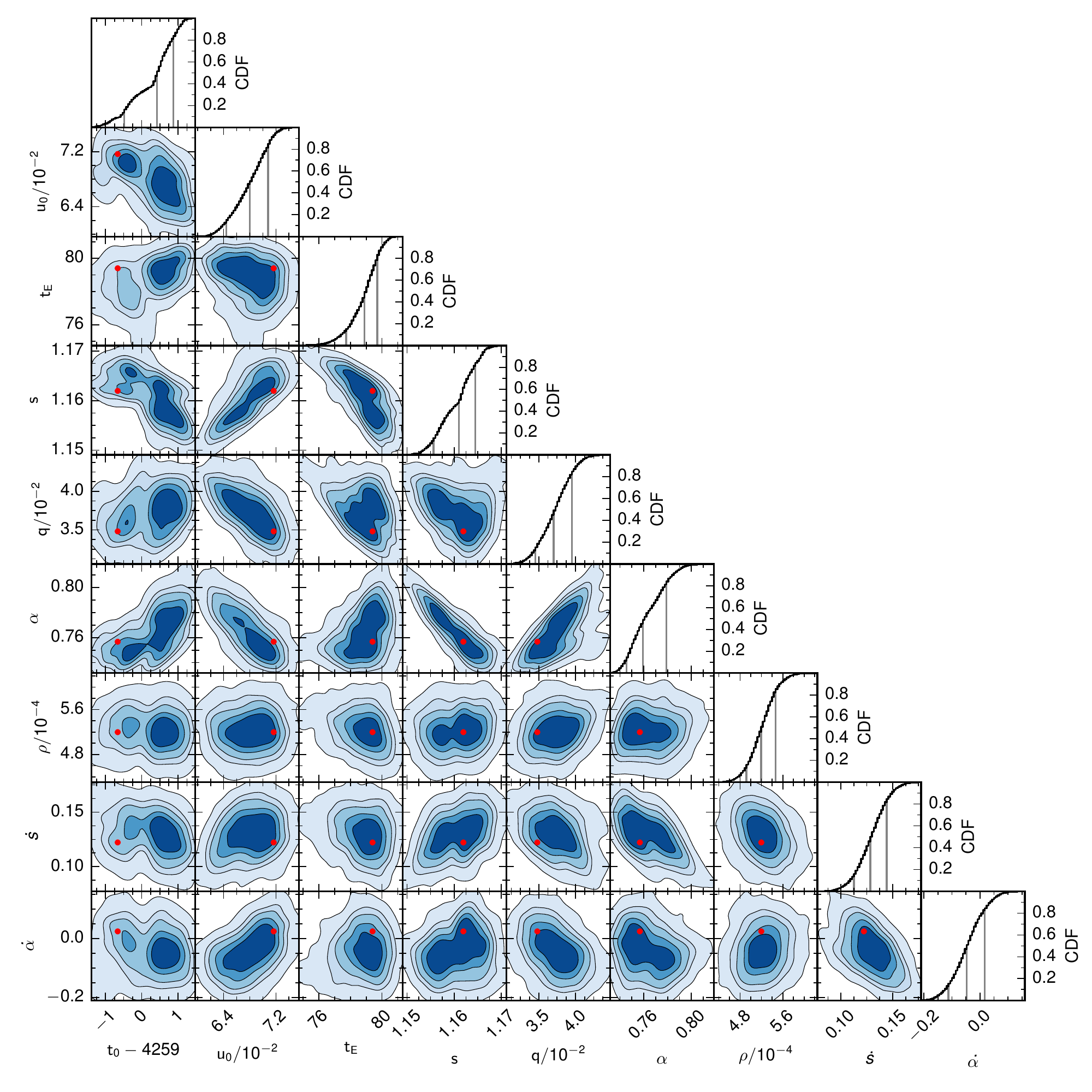}
\caption{Idem for ESBL+LOM's model.}
\label{fig:ESBLLOM}
\end{center}
\end{figure*}

	The best-fit model including lens orbital motion with ESBL is presented in column ESBL+LOM of \Tab{tab:lc-parameters} (correlation plots are shown in \Fig{fig:ESBLLOM}). According to this model, the changes in the caustic geometry are very slow, which should result in a poor signature on the light curve. Therefore, while this model also provides a better fit than ESBL alone ($\Delta\chi^2=93$), it is unlikely to give a good constraint on the lens orbital motion. Furthermore, the $\Delta\chi^2$ between this model and  ESBL+P (parallax alone) is only lower than 10 although two more parameters have been included in the model. We conclude that the key time intervals for disentangling orbital motion from parallax are not covered densely enough by the data. 

\subsection{Annual parallax and orbital motion}  \label{sec:lc-models-ESBL-P-OM}

	As (annual) parallax and orbital motion individually lead to a similar improvement of the ESBL $\chi^2$, we now include both effects in the model to check whether a better $\chi^2$ can be found. Parallax and orbital motion are known to be very correlated \citep{Batista2011,Skowron2011}. 
	
\begin{figure*}[ht]
\begin{center}
\includegraphics[width=\textwidth]{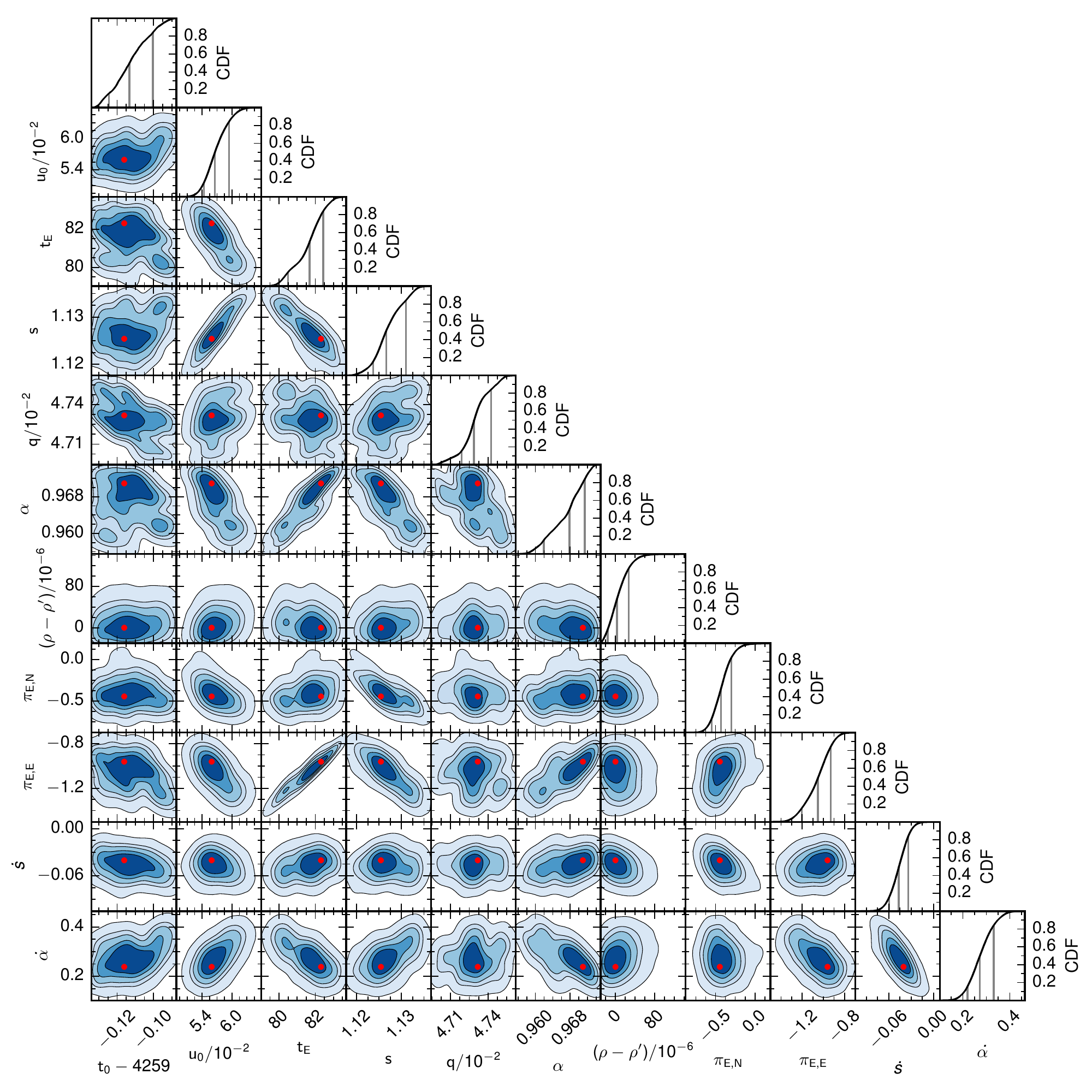}
\caption{Idem for ESBL+P+LOM's model ($u_0>0$) with $\rho' = 5.3\dix{-4}$.}
\label{fig:ESBLPLOM}
\end{center}
\end{figure*}

	We investigate the two cases $\uo>0$ and $\uo<0$ (columns ESBL+P+LOM of \Tab{tab:lc-parameters}) to check for possible ecliptic degeneracy \citep{Skowron2011}, and find that $\uo>0$ is preferred by $\Delta\chi^2=11.9$. The ESBL best-fit parameters, again, are relatively stable. The correlation plots of the best model are shown in \Fig{fig:ESBLPLOM}. As expected, the overall fit is not significantly better than ESBL+P ($\Delta\chi^2=12.2$) or ESBL+LOM alone ($\Delta\chi^2=20.5$). This is consistent with a strong degeneracy between these parameters, that cannot be reliably broken in the case of \MOA. 

\begin{table*}[ht]
\caption{Best-fit solutions for the different models of \MOA.}
\begin{center}
\begin{tabular}{l l l l l l}
\hline
\hline
Parameter \; [unit]                                               & \multicolumn{5}{c}{Model}\\\cline{2-6}
                                                                            & ESBL\tablefootmark{a}   & ESBL + P\tablefootmark{b}   & ESBL + LOM\tablefootmark{c}   & \multicolumn{2}{l}{\bfseries{ESBL + P + LOM}}\\
                                                                            &                                         &                                              &                                                    & $u_0<0$                     & $\mathbf{u_0>0}$\\
\hline
$\chi^2/\mbox{d.o.f.}$                                         & $(942.2/795)=1.19$         & $(840.9/792)=1.06$             & $(849.2/793)=1.07$                    & $(840.6/790)=1.06$    & $(828.7/790)=1.05$\\
$\Delta\chi^2=\chi^2-\chi^2_\mathrm{best}$       & $113.5$                           & $12.2$                                 & $20.5$                                         & $11.9$                         & $0$     \\
$s$                                                                      & $1.1291 \pm 0.0012$      & $1.1347\pm0.0020$            & $1.1620\pm0.0044$                    & $1.1361\pm0.0086$    & $1.1254\pm0.0044$\\
$q/10^{-2}$                                                          & $4.972 \pm 0.043$          & $4.867\pm0.080$                & $3.48\pm0.25$                           & $4.934\pm0.044$         & $4.732 \pm 0.020$\\
$\tE$ \; [days]                                                      & $80.72 \pm 0.21$            & $78.39\pm0.61$                  & $79.4\pm1.0$                             & $82.6\pm1.4$               & $82.3 \pm 1.2$\\
$\uo/10^{-2}$                                                       & $6.39 \pm 0.10$              & $5.57\pm0.15$                   & $7.17\pm0.32$                            & $-5.43\pm0.12$            & $5.59\pm 0.28$\\
$\to$ \; [THJD]                                                     & $4259.31 \pm 0.10$        & $4259.72\pm0.40$             & $4258.34\pm0.70$                      & $4259.23 \pm 0.10$      & $4258.884 \pm 0.013$\\
$\alpha/10^{-1}$ \; [rad]                                       & $9.123 \pm 0.056$          & $8.54\pm0.20$                   & $7.57\pm0.17$                            & $-9.62\pm0.14$             & $9.709 \pm 0.060$\\
$\rho/10^{-4}$   				                & $5.69 \pm 0.23$              & $5.91\pm0.30$                  & $5.20\pm0.27$                             & $6.13\pm0.33$              & $5.30\pm0.24$\\
$\piEN$                                                                & --            					      & $0.82\pm0.25$                  & --                                                   & $1.32 \pm 0.31$            & $-0.45 \pm 0.13$\\
$\piEE$                                                                & --              					      & $-0.31\pm0.15$                 & --                                                   & $0.82\pm0.65$             & $-0.96 \pm 0.15$\\
$\dot{s}/10^{-1}$ \;  [$\mathrm{rad/year}$]          & --                                     & --                                       & $1.22\pm0.16$                               & $-0.04\pm0.16$           & $-0.46 \pm 0.13$\\
$\dot{\alpha}/10^{-1}$ \; [$\mathrm{rad/year}$]   & --                                     & --                                       & $0.24\pm0.65$                               & $-2.76\pm0.60$           & $2.38 \pm 0.60$\\
\hline
\end{tabular}
\tablefoot{
\tablefoottext{a}{Extended-source binary-lens model.}
\tablefoottext{b}{Microlensing parallax.}
\tablefoottext{c}{Lens orbital motion.}
}
\label{tab:lc-parameters}
\end{center}
\end{table*}

\subsection{Summary and conclusions from the modelling} \label{sec:lc-discussion}

	A model with (annual) parallax improves the goodness-of-fit by $\Delta\chi^2 \sim 100$ compared to a static binary lens, but several models with very different values of $\piE$ give comparable values of $\chi^2$. Moreover, models with parallax or orbital motion alone lead to comparable $\chi^2$. This is not surprising for two reasons. Firstly, the caustic entry is not well covered by the data and it appears that a high value of parallax tends to change substantially the time of the caustic entry inside $E_1$. Secondly, parallax effects are partly degenerate with lens orbital motion, and the available data sets are not sufficient to disentangle these two effects.
	
We therefore adopt the ESBL parameters $(s,q,\uo, \alpha, \tE, \to)$ found for model ESBL+P+LOM and $u_0>0$.
Following the arguments summarised in the previous paragraph, other parameters such as parallax or orbital motion parameters cannot be used to constrain the lens mass and distance.

\begin{figure}[ht]
\centering
\includegraphics[scale=1]{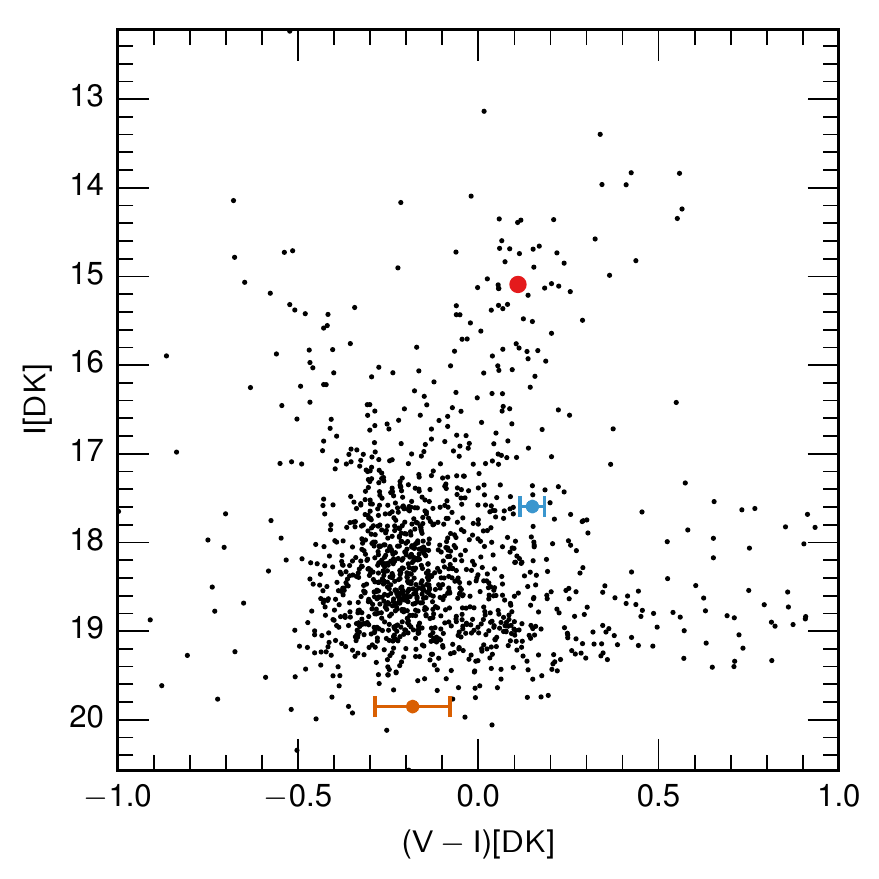}\\
\includegraphics[width=\linewidth]{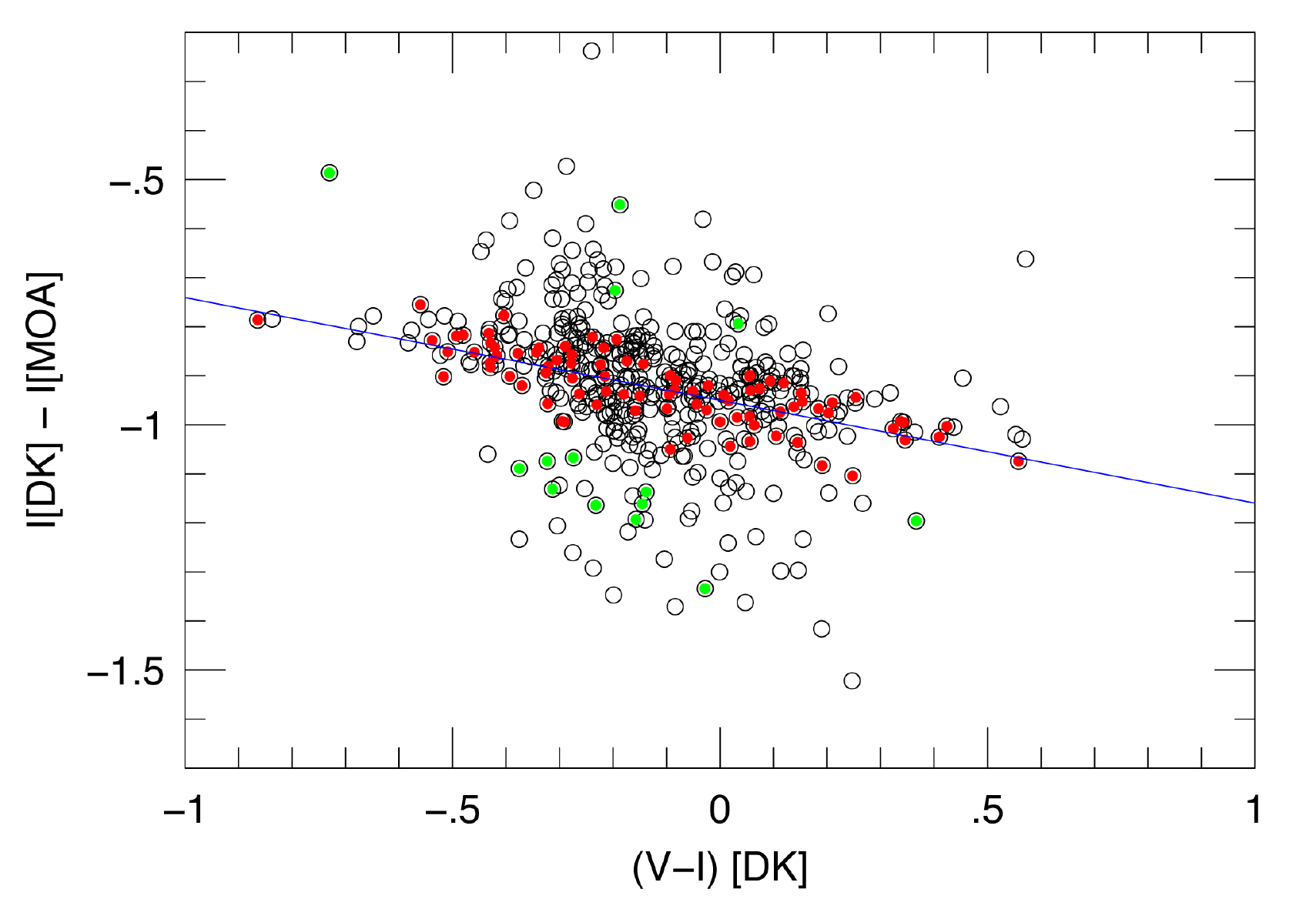}
\caption{Upper panel: $(I, V-I)$ Danish instrumental CMD of the field around \MOA, not corrected for interstellar extinction. The red point marks the position of the RCG. The magnitude of the microlensing target is in blue (\ie magnified source, lens and blended light). The orange point corresponds to the source alone. Lower panel: empirical linear relation between instrumental colours $(I_\mathrm{DK} - I_\mathrm{MOA})$ and $(V-I)_\mathrm{DK}$ (red and green points are astrometric matches that are uncrowded in the MOA image; after sigma-clipping, only red points are used for the fit).}
\label{fig:cmd_danish}
\end{figure}

\section{Physical parameters} \label{sec:physical-parameters-lens}
     
\subsection{Mass-distance relation from the source radius} \label{sec:physical-parameters-source}

	Combining \Eq{eq:rho} and \Eq{eq:thEpirel} yields the following mass-distance relation for the lens
\begin{equation}
	\ML = \frac{{\thS}^2}{\rho^2\kappa}\frac{1}{\pirel}\v
\label{eq:MDsource}
\end{equation}
where $\ML$ is the lens mass and $\pirel$ is a proxy to lens distance $\DL$ according to \Eq{eq:pirel}. In this equation, $\rho$ is measured from the light curve modelling (\Sec{sec:lc-discussion}) and $\thS$ is derived below.

	To characterise the source, we use a DoPHOT-based data reduction \citep{DoPhot} of Danish $I$ and $V$ data to build an instrumental $(I, (V-I))$ CMD (\Fig{fig:cmd_danish}). Fitting the $I$-band data for the best-fit model parameters provides $I_\mathrm{S,DK} =  19.85 \pm 0.01$. We use the RCG method to measure the interstellar extinction. Red clump giant stars are found to be relatively sparse, which makes it difficult to locate precisely the mean instrumental RCG position, $(I_\mathrm{RCG,DK}, (V-I)_\mathrm{RCG,DK})$. To overcome this problem, we cross-calibrate the Danish CMD with a CMD obtained at CTIO in a very similar $I$ filter, by cross-identifying a few clump stars. We then measure $I_\mathrm{RCG,DK}=15.09$ and  $(V-I)_\mathrm{RCG,DK}=0.11$. The values of $M_{I,\mathrm{RCG}}=-0.12 \pm 0.09$ and $(V-I)_\mathrm{RCG,0} = 1.06 \pm 0.12$ are taken from \cite{Nataf2013} at the coordinates of \MOA.
Measuring the dereddened apparent magnitude of the red clump stars located in the Galactic bulge, \cite{Nataf2013} found a distance to the Galactic centre $D_\mathrm{GC}=8.20~\mbox{kpc}$ and an angle between the Galactic bar and the line of sight from the Sun $\phi=40\degr$. From these values, the distance to the RCG is 
\begin{equation}
   D_\mathrm{RCG} = \frac{D_\mathrm{GC} \sin{\phi}}{\cos{(b)}\sin{(l+\phi)}}.
\end{equation}
For the Galactic coordinates of \MOA, the RCG is in the far side of the bar at a distance $D_\mathrm{RCG}=8.29~\mbox{kpc}$. We adopt a source at the same distance, corresponding to a distance modulus of $\mu=14.6 \pm 0.3\,\mathrm{mag}$, where the error bars also account for the uncertainty in the position of the RCG.

	We then calculate $I_\mathrm{S,0} = I_\mathrm{S,DK} + M_{I,\mathrm{RCG}} + \mu - I_\mathrm{RCG,DK} $ and obtain $I_\mathrm{S,0}=19.24 \pm 0.31$, which gives an absolute magnitude of $M_I=4.65\pm0.31$. We use a $10\,\mbox{Gyr}$ and solar metallicity isochrone from \cite{Bressan2012} to get the corresponding absolute magnitude $M_V$  of the source
from which we derive an intrinsic source colour of $(V-I)_\mathrm{S,0}=0.88 \pm 0.1$. This value can be compared to an independent estimate of the source colour based on the method of \cite{Gould2010}. MOA $R_M$-band images are reduced with DoPHOT, and stars are cross-matched between Danish $I$, $V$ and MOA $R_M$ frames (lower panel of  \Fig{fig:cmd_danish}). We obtain
\begin{equation}
	I_\mathrm{S,DK} - I_\mathrm{S,MOA} = -(0.950 \pm 0.006) - (0.21 \pm 0.02) \times(V-I)_\mathrm{S,DK}\v
\label{eq:methgou}
\end{equation}
where  $I_\mathrm{S,MOA} = 20.76 \pm 0.02$ from the fit of the $R_M$ light curve using the best-fit parameters. Hence, $I_\mathrm{S,DK} - I_\mathrm{S,MOA} = -0.91 \pm 0.02$, which yields $(V-I)_\mathrm{S,DK} = -0.18 \pm 0.10$. It follows that the estimated de-reddened source colour is $(V-I)_\mathrm{S,0} = (V-I)_\mathrm{S,DK} + (V-I)_\mathrm{RCG}-(V-I)_\mathrm{RCG,DK}=0.77 \pm 0.2$, which is in good agreement with the previous estimate.  In the following, we adopt $(V-I)_\mathrm{S,0}=0.88 \pm 0.1$ (most robust estimate) and $I_\mathrm{S,0}=19.24 \pm 0.31$. The source is thus a G6-K0 Main Sequence star.

	From \cite{Kervella2008} brightness-colour relations, we estimate the angular radius of the source,
\begin{equation}
	\log(\thS) = 3.198 - 0.2 I_\mathrm{S,0} + 0.4895 (V-I)_\mathrm{S,0}- 0.0657 (V-I)_\mathrm{S,0}^2\v
\end{equation}
which yields $\thS = 0.54 \pm 0.05~\mbox{$\mu$as}$. With a source at $\DS=8.3$ kpc, the physical source radius is $\RS=\DS\thS=0.96\ R_\sun$. From \cite{ModernAstro2006book}, this radius is that of G8 Main Sequence star, which is in very good agreement with the constraint from the colour. 

	The lens mass-distance relation obtained from \Eq{eq:MDsource} where $\thS = 0.54 ~\mbox{$\mu$as}$ and  $\rho=5.30\dix{-4}$ is plotted in \Fig{fig:correlation_post}.

\subsection{Mass-distance relation from NaCo data} \label{sec:NACOmassdist}

\begin{figure}[ht]
\begin{center}
\includegraphics[scale=1]{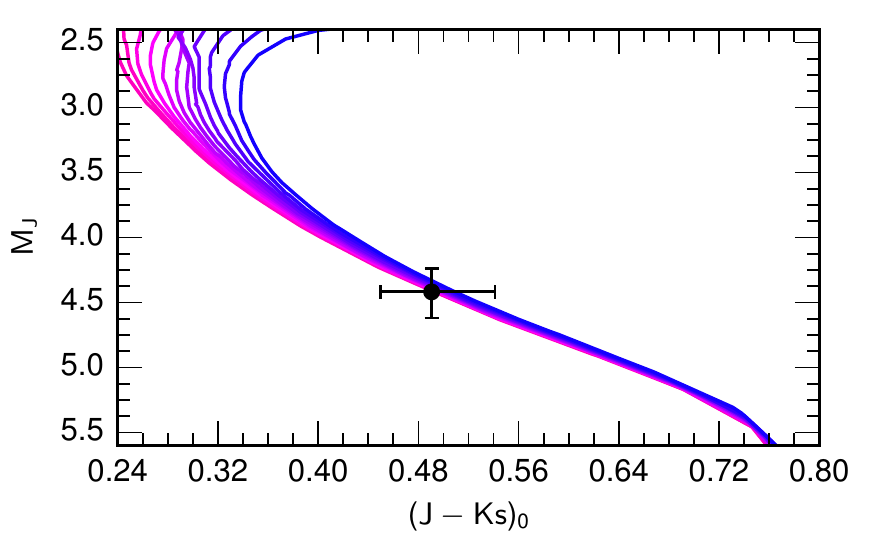}
\caption{Colour-magnitude diagram $(M_J,(J-K_s)_0)$ of the lens. The curves are a set of isochrones from $2~\mbox{Gyr}$ (in pink) to $8~\mbox{Gyr}$ (in blue) with solar metallicity \citep{Bressan2012}. The black cross indicates the range of colours and magnitudes explored by the MCMC at the 1-$\sigma$ level.}
\label{fig:cmdjklens}
\end{center}
\end{figure}
 
	Calibrated ($\magJ, \magH,\magKs$) NaCo magnitudes (\Sec{sec:datanaco}) provide independently from \Eq{eq:MDsource} further lens mass-distance relations (written below for $\magJ$ only) through 
\begin{equation}
 	m_\magJ(L) = M_\magJ + 5\log\DL -5 + A_\magJ \v
\end{equation}
where $m_\magJ(L)$ is the lens apparent reddened magnitude of the lens, $A_\magJ$ the interstellar absorption (given in \Sec{sec:datanaco}), and $M_\magJ$ the lens absolute magnitude (which is a function of its mass $\ML$). The source ($\magJ, \magH,\magKs$) magnitudes are expressed as (e.g.,  for $\magJ$)
\begin{equation}
 	m_\magJ(S) = M_\magJ + 5\log\DS -5 + A_\magJ  - 2.5\log(A)\v
\end{equation}
where $A=1.3$ is the magnification of the source at the time of the NaCo observations, and assuming that the absorption mostly occurs in the first kiloparsecs towards the bulge.
The resulting ($\magJ, \magH,\magKs$) magnitudes are then computed from the previous equations.

	The absolute magnitudes $M_{\magJ, \magH,\magKs}$ are computed using the isochrones by \cite{Bressan2012} for both the source and lens. For the source we use the same isochrone as in \Sec{sec:physical-parameters-source}.
	For the lens, we assume an age of $2.5~\mbox{Gyr}$ and solar metallicity. In fact, as seen in \Fig{fig:cmdjklens}, the lens lies in a region of the isochrones where the age has only very small impact on the magnitudes. The choice of metallicity has slightly more impact, but keeps the variation lower than the magnitude error bars for metallicity changes as large as $Z_\sun\pm0.02$. 

\begin{figure}[t]
\centering
\includegraphics[scale=1]{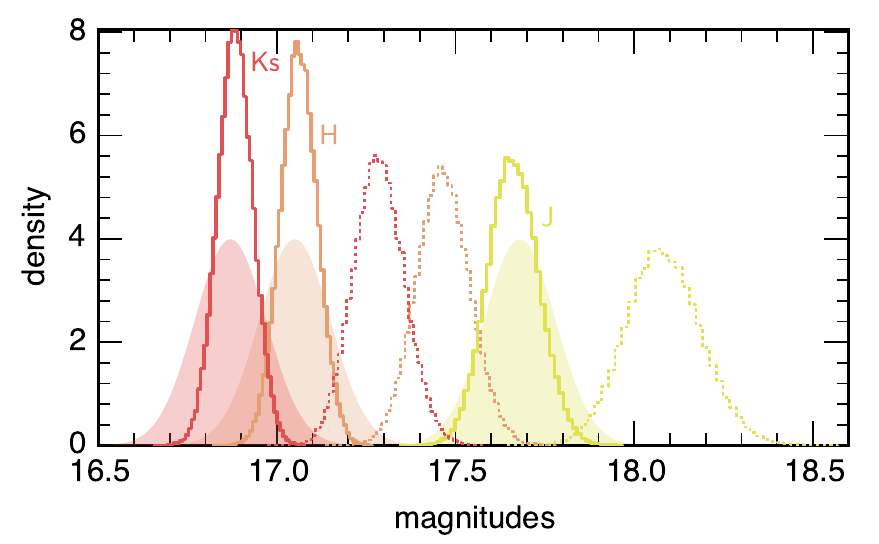}
\caption{Probability densities of NaCo  ($\magJ, \magH,\magKs$) magnitudes, not corrected for interstellar extinction. Filled curves are Bayesian Gaussian priors based on the NaCo measurements, which are compared to posterior distributions (solid lines). The dotted posterior distributions are those of the lens magnitudes alone (\ie, disentangled from the source; respectivement $\magJ,  \magH, \magKs$ from the right to the left).}
\label{fig:posterior}
\end{figure}

	We then use a MCMC algorithm (\Sec{sec:MDcomb}) to match the observed NaCo magnitudes listed in \Tab{tab:NaComags} (Epoch 1) to the theoretical magnitudes, using the lens mass $\ML$ and distance $\DL$ as fitting parameters. The observed magnitudes serve as Gaussian prior distributions with  standard deviation equal to the magnitudes error bars; they are shown as filled curves in \Fig{fig:posterior}. In the same figure, posterior distributions for the resulting magnitudes (lens, and source+lens) are shown in solid and dashed lines respectively. The final lens mass-distance relations are plotted in \Fig{fig:correlation_post}. It can be seen that while in principle three relations are obtained, they are very strongly correlated and thus only correspond to one effective independent mass-distance relation.

\subsection{Mass-distance relation from parallax or orbital motion}

	In principle, when parallax is measured, a further independent lens mass-distance relation can be obtained by combining \Eq{eq:piE} with \Eq{eq:thEpirel}, 
\begin{equation}
	\ML = \frac{\pirel}{\piE^2\kappa}\v
\label{eq:MqLePara}
\end{equation}
where $\piE$ is derived from the light curve modelling. Nevertheless, as discussed in \Sec{sec:lc-discussion},  the parallax is degenerate with the lens orbital motion, which prevent us from using this constraint. We however plot on \Fig{fig:correlation_post} the typical behaviour of the mass-distance relation obtained from parallax measurements, if $\piE$  had been measured. We discuss this problem further below. 

\subsection{Combination of the constraints and physical parameters} \label{sec:MDcomb}

\begin{figure}[!t]
\centering
\includegraphics[scale=1]{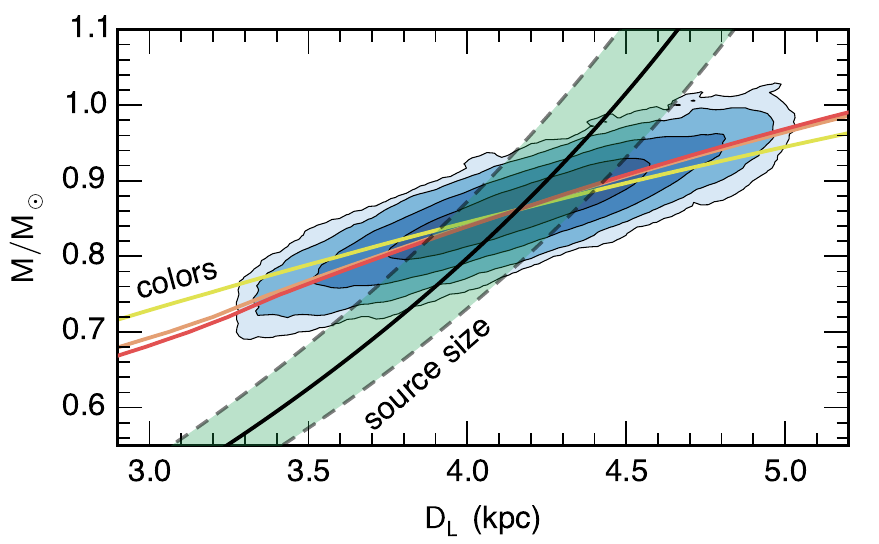}
\caption{Lens mass-distance relations derived from the source size $\rho$ (1-$\sigma$ around the best value shown by the green shadow limited by the dashed lines) and NaCo ($\magJ, \magH,\magKs$) colours constrains (respectively yellow, orange, red from top to bottom below the label ``colours'').
The blue contours ($1-4\,\sigma$) represent the joint posterior probability density $P(\ML,\DL)$ from the MCMC run.}
\label{fig:correlation_post}
\end{figure}

	We combined the different mass-distance relations discussed in the previous sections (and shown in \Fig{fig:correlation_post}) in a MCMC minimisation process to recover the lens mass $\ML$ and the observer-lens distance $\DL$. We use the measurement and error bars for $\rho$, ($\magJ, \magH,\magKs$) and $\piE$ to build up Gaussian priors for the Bayesian analysis, and flat (uninformative) priors for $\ML$ and $\DL$. The values of the parameters correspond to the solution $u_0>0$, as discussed in \Sec{sec:lc-discussion}. We used the convergence criterion of \cite{Geweke1992} to stop the MCMC and compute the posterior probability densities. 
	
	As seen in \Fig{fig:correlation_post}, in principle  this problem is over-constrained, with three relations for only two fitting parameters. This enabled us to check the consistency of the different measurements, and investigate further the degeneracy between parallax and lens orbital motion, which prevents $\piE$ from being used as a constraint. We first noticed a clear discrepancy between the value of $\piE$ derived from the fit (the prior) and the posterior value. This pointed out a strong tension between the three constraints from parallax, source size and infrared colours. We then successively removed the parallax, source size or colours constraints from the MCMC runs. Without the source size constraint, a model with parallax values as large as $\piE \sim 1.1$ was found. This would require a much closer lens ($\DL\sim0.6\,\mbox{kpc}$), a longer Einstein radius ($\thE\sim1.6\, \mas$) and consequently a much smaller source size ($\rho\sim3.3\dix{-4}$). This value of $\rho$ is many sigma away from that measured from the light curve modelling. Such a difference  is very unlikely given the very strong constraint on $\rho$ obtained from the caustic exit modelling. This therefore confirmed previous concerns that in this case, the degeneracy between parallax and lens orbital motion does not allow us to use the parallax mass-distance relation. Consequently, the parallax constraints are removed from the MCMC to derived the final values for $\ML$ and $\DL$, thus constrained by the source size and the colours.

	The final values for $\ML$ and $\DL$ are given in \Tab{tab:final-model}, with the correlation plot shown in \Fig{fig:correlation_post}. We also compute the mass of the lens companion, $\MP = q\ML$, and find $\MP = 41 \pm 2\,\MJ$. The host mass is $0.82 \pm 0.04 \, \Msun$. The companion is thus a brown dwarf orbiting a solar-type star.
		
\begin{table}[ht]
\begin{center}
\begin{tabular}{ll}
\hline
\hline
Physical parameter [unit] & Value \\
\hline
Lens mass, $\ML$ [$\Msun$] & $0.86 \pm 0.04$ \\
Host mass [$\Msun$] & $0.82 \pm 0.04$ \\
Companion mass, $\MP$ [$\MJ$] & $41 \pm 2$ \\ 
Projected orbit, $a_\perp$ [AU] & $4.29 \pm 0.10$ \\
Observer-lens distance, $\DL$ [kpc] & $4.17 \pm 0.30$ \\
\hline
$\DS$ [kpc] & $8.3 \pm 1.0$ \\
$\thS$ [$\mu\mathrm{as}]$ & $0.54\pm 0.05$ \\
$\thE$ [mas] & $0.91 \pm 0.04$ \\
$\vperp$ [$\mbox{km.s}^{-1}$] & $80\pm2$ \\
$\piE$ (calculated) [mas] & $0.13 \pm 0.01$ \\ 
$\gamma$ (new fit) [year$^{-1}$] & $0.26 \pm 0.07$ \\
$\beta$ & $0.08 \pm 0.07$\\
 \hline
\end{tabular}
\end{center}
\caption{Physical parameters of \MOAL and its companion.}
\label{tab:final-model}
\end{table}

	We finally perform additional consistency checks on the overall parameters of \MOAL. First, we compute the projected lens-source relative velocity
\begin{equation}	
	\vperp = \frac{\thE\DL}{\tE}\v
\end{equation}
and find $\vperp \simeq 80\,\mbox{km.s}^{-1}$, which is in very good agreement with  probability densities predicted by \cite{Dominik2006} for the derived values of $\tE$ and $\DL$.
Moreover and as a supplementary check, we computed the probability distribution of $\piE$ from the distributions of $\ML$ and $\DL$ using \Eq{eq:MqLePara}, and derived its maximum a posteriori (MAP) value. We then ran again the light curve modelling with this value of $\piE$ kept fixed. For the overall consistency, all the parameters are fixed as well, except $\dot{s}$ and $\dot{\alpha}$ (values are given in \Tab{tab:lc-parameters}, model with $u_0>0$) which yields the value of the orbital motion parameter $\gamma$ (see \Tab{tab:final-model}).
From this, we compute 
\begin{equation}
   \beta = \left|\frac{\Ecin}{\Epot}\right| = \frac{2\mbox{AU}^2}{c^2} \frac{\piE\, s^3\gamma^2}{\thE\left(\piE + \piS/\thE \right)^3} \v
\end{equation}
the ratio of the apparent kinetic to potential energy for the binary lens orbit projected onto the plane of the sky \citep{An2002}. Gravitationally bound systems should exhibit $0<\beta<1$ (for face-on circular orbits, $\beta=0.5$). Using $\piS=\mbox{AU}/\DS$, we find $\beta \simeq 0.08$, which is a value consistent with a bound system. In contrast, high parallax values of e.g., $\piE=1.1$ lead to much lower values of $\beta\simeq 4.5\dix{-3}$. As argued by \cite{Batista2011}, a scenario with $\beta \ll 1$ would require highly improbable orbital parameters or projections, such as a very close-in companion seen on a  nearly edge-on circular orbit, which is excluded here.

\section{\MOALb in the brown dwarf landscape} \label{sec:perspectives}

	While around $1900$ exoplanets have been detected so far\footnote{\url{http://exoplanet.eu}}, less than a hundred brown dwarfs orbiting stars are known today. The lack of brown dwarf companions to solar-type stars was noticed early after the first exoplanets detections, and referred to as a ``brown dwarf desert''. Using the radial velocity method, \cite{Marcy2000} noted that at tight orbital separations ($a<3\au$), companions more massive than $8~\MJ$ represented only a very low fraction of the detected objects. This was not expected from the relatively high frequency ($\sim13\%$) of binary stars found earlier by \cite{Duquennoy1991}, also at close separation. 
	
	This lack of brown dwarf companions compared to planetary companions has since been confirmed by all detection methods. For instance, \cite{Lafreniere2007} found a frequency of $1.9^{+8.3}_{-1.5}\%$ for  $13-75~\MJ$ companions located in the range $25-250~\au$ from the Gemini Deep Planet Survey. Similarly, \cite{Metchev2009} derived a frequency of $3.2\pm3.1\%$ for brown dwarfs orbiting young solar-type stars in the range $28-1590~\au$ using adaptive optics direct imaging. \cite{McCarthy2004} also found a higher proportion of stellar compared to substellar companions to stars, and derived a frequency ratio of $\sim 3-10$ at wide separations. \cite{Luhman2007} found that this ratio is comparable to the relative abundance of stars to substellar objects when they are found single, either in star forming regions or in the solar neighbourhood. 

\begin{figure}[t]
\begin{center}
\includegraphics[scale=1]{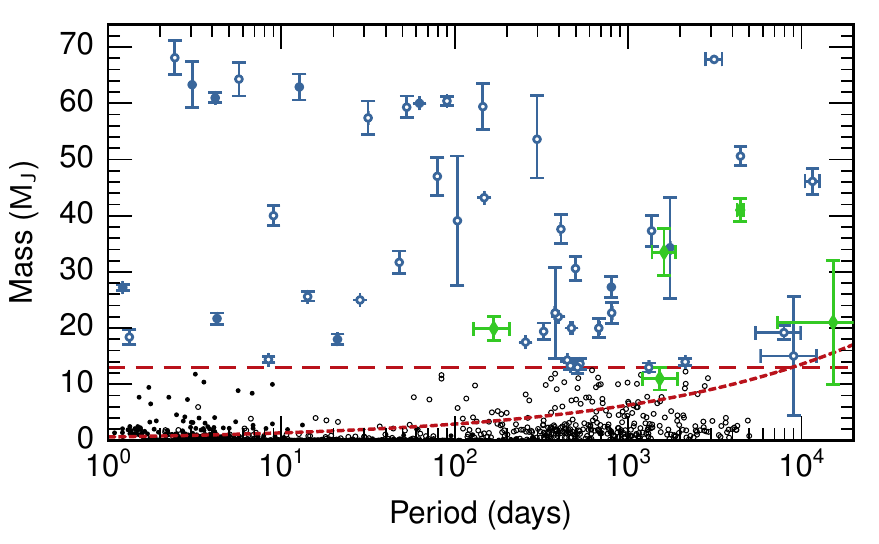}
\includegraphics[scale=1]{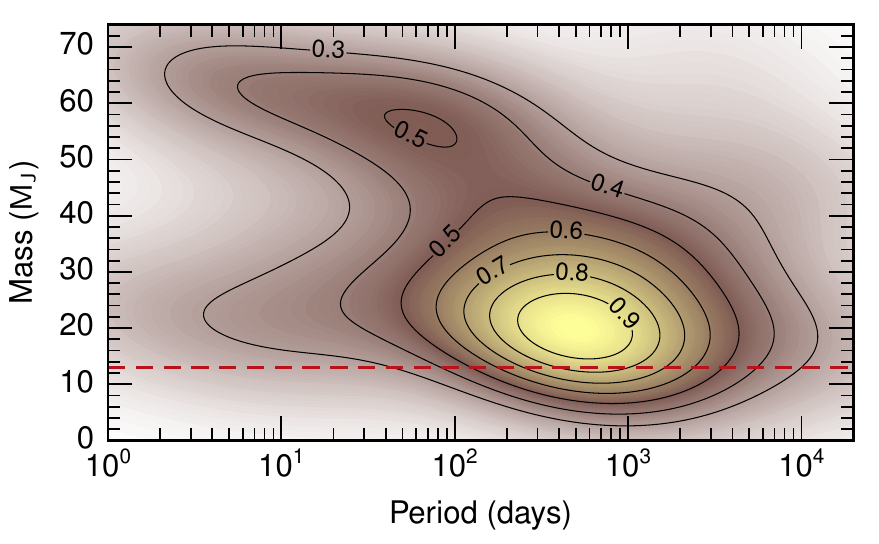}
\caption{The upper panel shows a mass-period diagram with brown dwarfs companions detected through radial velocity, transit and direct imaging \citep[blue filled circles for measured masses, and open circles for minimum masses; list adapted from][]{Ma2014}, and microlensing (green diamonds). For reference, exoplanets detected so far are also displayed (small black circles, \url{http://exoplanets.org}). The red dotted line indicates the global radial velocity completeness limit, while the red dashed line marks the region above which data are included to perform the non-parametric, two-dimensional probability density distribution shown in the lower panel.}
\label{fig:massperiod}
\end{center}
\end{figure}

	The upper panel of \Fig{fig:massperiod} is a mass vs. period diagram which diplays the known brown dwarf companions of solar-type stars detected through microlensing, radial velocimetry, transit and direct imaging. For radial velocimetry and transit, we included objects from \cite{Ma2014} catalogue with (minimum) masses in the range $13-74~\MJ$ and orbital periods below $2\dix{4}\,\mathrm{d}$. Furthermore, we have excluded objects with mass uncertainties above  $25\,\MJ$, and deleted (false positive) TYC~1240-945-1.
Microlensing brown dwarfs are shown as green points, they are: MOA-2011-BLG-149 and OGLE-2011-BLG-0172/MOA-2011-BLG-104 \citep{Shin2012b}, MOA-2010-BLG-073 \citep{Street2013}, OGLE-2013-BLG-0578 \citep{Park2015arXiv} as well as \MOA (this paper). For consistency with Doppler and transit data, we did not include MOA-2009-BLG-411 \citep{Bachelet2012b} which has a large uncertainty in the mass. In the case of microlensing objects for which only projected separations $a_\perp$ were measured (all objects in this case), we estimate the (physical) semi-major axis $a$ as the median of the probability distribution of  \cite{Gould1992}, which we find to be equal to $a = (2/\sqrt{3}) a_\perp\approx 1.15a_\perp$. Kepler's third law is then used to yield the estimated period. The error bar on measured $a_\perp$ is propagated to $a$ following the previous equation, and the error bar on the period is obtained through a MCMC run including uncertainties on $a$ and primary mass $M$ (Gaussian distributions are assumed).

	As seen in \Fig{fig:massperiod}, the distribution of brown dwarfs is not uniform, in particular it exhibits an increasing frequency of objects with increasing orbital period. Furthermore, \cite{Ma2014} argue that the driest part of the brown dwarf desert seems to be confined within a region at close separation, namely within $P<100\,\mathrm{d}$ and for masses within $30-55~\MJ$ for Sun-like host stars. They subsequently split the objects into two distinct populations, whether their mass is higher or lower than $42.5\,\MJ$: the more massive brown dwarfs would mainly be the outcome of gravitational fragmentation and collapse of a molecular cloud (star-like formation scenario, corroborated by an eccentricity distribution similar to binary stars), while brown dwarfs below $42.5\,\MJ$ would mainly result from gravitational instability within the proto-planetary disk. But other authors such as  \cite{Guillot2012}, however, argue that the depletion of objects at tight orbits may as well be explained by a loss of an initial population of close-in brown dwarfs due to tidal interactions with their host stars. In this scenario, close-in, massive objects lose angular momentum due to the slower rotation of the star relative to the planets orbital motion, spiral in and fall into the star. This effect is predicted to peak for G-dwarf primaries and should not be important for earlier-type stars, which is supported by the detection of close-in brown dwarfs around F-type stars.  

	Different mechanisms may overlap to shape the brown dwarfs landscape, and lead to a more complex structure than previously discussed. To analyse it further, we performed a non-parametric, two-dimensional probability density distribution fit to the data, only for objects above the red dashed line (\ie, above the radial velocity completeness limit, based on \cite{Mayor2011}, and marked by the red dotted line in \Fig{fig:massperiod}). The probability density estimation is based on \cite{Scott1992}; in practice we used a Gaussian kernel, and follow \cite{Silverman1986} rule to estimate the bandwidth parameter. The resulting probability density is shown in the lower panel of \Fig{fig:massperiod}.
We used two different methods to compute the bandwidth parameter,  \cite{Silverman1986} or \cite{Scott1992}, and found that the resulting probability density is relatively stable, even if a few data points are removed from sparsely sampled regions.
	
	The first striking feature of the density profile (lower panel of \Fig{fig:massperiod}) is a region of depletion of objects at intermediate masses ($M \sim 30-60\,\MJ$) and short orbital periods ($P\lesssim 30\,\mathrm{d}$). This matches the region referred to as the driest part of the brown dwarf desert according to \cite{Ma2014}. Second, an accumulation of objects can be seen around $P \sim 500\,\mathrm{d}$ and $M \sim 20\,\MJ$ (following the apparent trend observed for giant planets, black data points in \Fig{fig:massperiod}). Third, we find another depletion of objects at long orbital periods ($P \gtrsim 500\,\mathrm{d}$) and high masses ($M \gtrsim 50\,\MJ$).
	Since the brown dwarf sample is drawn from various surveys (mainly from radial velocities), the interpretation of these features should be taken with caution. Although all brown dwarfs in \Fig{fig:massperiod} are chosen above the radial velocity completeness limit from \cite{Mayor2011}, it is not guaranteed that these objects are not affected by observational biases. In particular, a degeneracy between mass and period is increasingly affecting very long-period brown dwarfs, and the different surveys should be corrected from their sample size. Hence, while a thorough analysis of all these factors is necessary to assess the exact shape of the brown dwarfs distribution, the gross features emerging from the lower panel of \Fig{fig:massperiod} may well be real. 
	If we split the density profile into two regions of masses above and below $42.5\,\MJ$ as discussed by \cite{Ma2014}, the distribution of high-mass objects can be seen as shifted towards  shorts orbital periods ($P \sim 30\,\mathrm{d}$), while less massive objects appear to pile up at longer orbital periods  ($P \sim 500\,\mathrm{d}$). This supports the claim of  \cite{Ma2014} that massive objects would accumulate at short periods as a result of gravitational collapse of a molecular cloud, contrary to less-massive objects built up by gravitational instability in the disk which would accumulate at longer periods. Nevertheless, the depletion at intermediate masses and short periods would as well be affected by a star-engulfing mechanism advocated by \cite{Guillot2012}, in particular because almost all brown dwarfs included here orbit Sun-like stars, for which this effect peaks.

	The distribution of brown dwarfs as a function of mass and period still remains uncertain because of a lack of detections. It is thus difficult to distinguish between the different mechanisms which shape the brown dwarf desert. Moreover, these detections are mostly objects orbiting Sun-like stars, which makes it difficult to study their distribution as a function of host star mass. While the \MOAL companion orbits a G-K dwarf host, microlensing hosts are most frequently M-K dwarfs. Hence, future microlensing surveys will provide unique information on the brown dwarf distribution for low-mass hosts, thus offering a complementary mass bin to other methods. 

\section{Summary and prospects}

	We have presented \MOALb, the first brown dwarf discovered around a Sun-like star through gravitational microlensing. The system is located at $4.2\pm0.3\,\kpc{}$ from the Earth, the brown dwarf companion has a mass of $41\pm 2\,\MJ$ and was observed  at a projected distance of $4.3\pm0.1\,\au$ from a K dwarf host star. We have performed a non-parametric probability density distribution fit to the population of brown dwarf companions detected so far. The resulting density profile has a complex structure, leading to a ‘‘brown dwarf landscape’’ most likely shaped by different (and perhaps competitive) formation or destruction mechanisms. While it seems difficult with the current data set to distinguish observationally which are the dominant mechanisms, an answer to this question appears to within the reach of further observations in the near to mid-term future.
	
	 Gravitational microlensing is an exceptional tool to detect brown dwarfs as free-floating objects, companions to stars or as brown dwarfs binaries. It has a unique sensitivity to detect brown dwarfs companions to stars of any type, in particular at long orbital periods. Recent advances in using networks of robotic telescopes \citep[e.g.,][]{Tsapras2009} will provide in a near future an order of magnitudes more brown dwarfs detections through microlensing. Future microlensing space missions \citep{PennyEuclid,Yee2014Exopag,Euclid2010} also carry important promises for providing unique information on the populations of brown dwarfs in their different configurations.

\begin{acknowledgements}
A.C. acknowledges financial support from the Emergence UPMC 2012 grant. C.R. and A.C. are grateful to Bo Ma for providing us with the brown dwarfs catalogue used in \cite{Ma2014}. We especially thank the University of Tasmania for granting us access to their TPAC supercomputer facilities where part of the calculations were carried out. Work by C.H. was supported by Creative Research Initiative Program (2009-0081561) of National Research Foundation of Korea. K.H., M.D. and M.H. are supported by NPRP grant NPRP-09-476-1-78 from the Qatar National Research Fund (a member of Qatar Foundation). M.H. acknowledges support from the Villum foundation. S.D. is supported by ``the Strategic Priority Research Program- The
Emergence of Cosmological Structures'' of the Chinese Academy of Sciences (grant No. XDB09000000).
\end{acknowledgements}

\bibliographystyle{aa}
\bibliography{references}

\begin{thebibliography}{77}
\expandafter\ifx\csname natexlab\endcsname\relax\def\natexlab#1{#1}\fi

\bibitem[{{Alard}(2000)}]{ISIS}
{Alard}, C. 2000, \aaps, 144, 363

\bibitem[{{Alard} \& {Lupton}(1998)}]{AlardLupton}
{Alard}, C. \& {Lupton}, R.~H. 1998, \apj, 503, 325

\bibitem[{{Albrow} {et~al.}(1999{\natexlab{a}}){Albrow}, {Beaulieu},
  {Caldwell}, {Depoy}, {Dominik}, {Gaudi}, {Gould}, {Greenhill}, {Hill},
  {Kane}, {Martin}, {Menzies}, {Naber}, {Pogge}, {Pollard}, {Sackett}, {Sahu},
  {Vermaak}, {Watson}, {Williams}, \& {The PLANET Collaboration}}]{Albrow1999}
{Albrow}, M.~D., {Beaulieu}, J.-P., {Caldwell}, J.~A.~R., {et~al.}
  1999{\natexlab{a}}, \apj, 522, 1022

\bibitem[{{Albrow} {et~al.}(1999{\natexlab{b}}){Albrow}, {Beaulieu},
  {Caldwell}, {Dominik}, {Greenhill}, {Hill}, {Kane}, {Martin}, {Menzies},
  {Naber}, {Pel}, {Pollard}, {Sackett}, {Sahu}, {Vermaak}, {Watson},
  {Williams}, {Sahu}, \& {PLANET Collaboration}}]{Albrow1999LD}
{Albrow}, M.~D., {Beaulieu}, J.-P., {Caldwell}, J.~A.~R., {et~al.}
  1999{\natexlab{b}}, \apj, 522, 1011

\bibitem[{{Albrow} {et~al.}(2009){Albrow}, {Horne}, {Bramich}, {Fouqu{\'e}},
  {Miller}, {Beaulieu}, {Coutures}, {Menzies}, {Williams}, {Batista},
  {Bennett}, {Brillant}, {Cassan}, {Dieters}, {Prester}, {Donatowicz},
  {Greenhill}, {Kains}, {Kane}, {Kubas}, {Marquette}, {Pollard}, {Sahu},
  {Tsapras}, {Wambsganss}, \& {Zub}}]{Albrow2009}
{Albrow}, M.~D., {Horne}, K., {Bramich}, D.~M., {et~al.} 2009, \mnras, 397,
  2099

\bibitem[{{An} {et~al.}(2002){An}, {Albrow}, {Beaulieu}, {Caldwell}, {DePoy},
  {Dominik}, {Gaudi}, {Gould}, {Greenhill}, {Hill}, {Kane}, {Martin},
  {Menzies}, {Pogge}, {Pollard}, {Sackett}, {Sahu}, {Vermaak}, {Watson}, \&
  {Williams}}]{An2002}
{An}, J.~H., {Albrow}, M.~D., {Beaulieu}, J.-P., {et~al.} 2002, \apj, 572, 521

\bibitem[{{Bachelet} {et~al.}(2012){Bachelet}, {Fouqu{\'e}}, {Han}, {Gould},
  {Albrow}, {Beaulieu}, {Bertin}, {Bond}, {Christie}, {Heyrovsk{\'y}}, {Horne},
  {J{\o}rgensen}, {Maoz}, {Mathiasen}, {Matsunaga}, {McCormick}, {Menzies},
  {Nataf}, {Natusch}, {Oi}, {Renon}, {Tsapras}, {Udalski}, {Yee}, {Batista},
  {Bennett}, {Brillant}, {Caldwell}, {Cassan}, {Cole}, {Cook}, {Coutures},
  {Dieters}, {Dominik}, {Dominis Prester}, {Donatowicz}, {Greenhill}, {Kains},
  {Kane}, {Marquette}, {Martin}, {Pollard}, {Sahu}, {Street}, {Wambsganss},
  {Williams}, {Zub}, {PLANET Collaboration}, {Bos}, {Dong}, {Drummond},
  {Gaudi}, {Graff}, {Janczak}, {Kaspi}, {Koz{\l}owski}, {Lee}, {Monard},
  {Mu{\~n}oz}, {Park}, {Pogge}, {Polishook}, {Shporer}, {Fun Collaboration},
  {Abe}, {Botzler}, {Fukui}, {Furusawa}, {Hearnshaw}, {Itow}, {Korpela},
  {Ling}, {Masuda}, {Matsubara}, {Miyake}, {Muraki}, {Ohnishi}, {Rattenbury},
  {Saito}, {Sullivan}, {Sumi}, {Suzuki}, {Sweatman}, {Tristram}, {Wada}, {Moa
  Collaboration}, {Allan}, {Bode}, {Bramich}, {Clay}, {Fraser}, {Hawkins},
  {Kerins}, {Lister}, {Mottram}, {Saunders}, {Snodgrass}, {Steele}, {Wheatley},
  {Robonet-Ii Collaboration}, {Bozza}, {Browne}, {Burgdorf}, {Calchi Novati},
  {Dreizler}, {Finet}, {Glitrup}, {Grundahl}, {Harps{\o}E}, {Hessman}, {Hinse},
  {Hundertmark}, {Liebig}, {Maier}, {Mancini}, {Rahvar}, {Ricci}, {Scarpetta},
  {Skottfelt}, {Southworth}, {Surdej}, {Zimmer}, \& {Mindstep
  Consortium}}]{Bachelet2012b}
{Bachelet}, E., {Fouqu{\'e}}, P., {Han}, C., {et~al.} 2012, \aap, 547, A55

\bibitem[{{Batista} {et~al.}(2011){Batista}, {Gould}, {Dieters}, {Dong},
  {Bond}, {Beaulieu}, {Maoz}, {Monard}, {Christie}, {McCormick}, {Albrow},
  {Horne}, {Tsapras}, {Burgdorf}, {Calchi Novati}, {Skottfelt}, {Caldwell},
  {Koz{\l}owski}, {Kubas}, {Gaudi}, {Han}, {Bennett}, {An}, {MOA
  Collaboration}, {Abe}, {Botzler}, {Douchin}, {Freeman}, {Fukui}, {Furusawa},
  {Hearnshaw}, {Hosaka}, {Itow}, {Kamiya}, {Kilmartin}, {Korpela}, {Lin},
  {Ling}, {Makita}, {Masuda}, {Matsubara}, {Miyake}, {Muraki}, {Nagaya},
  {Nishimoto}, {Ohnishi}, {Okumura}, {Perrott}, {Rattenbury}, {Saito},
  {Sullivan}, {Sumi}, {Sweatman}, {Tristram}, {von Seggern}, {Yock}, {PLANET
  Collaboration}, {Brillant}, {Calitz}, {Cassan}, {Cole}, {Cook}, {Coutures},
  {Dominis Prester}, {Donatowicz}, {Greenhill}, {Hoffman}, {Jablonski}, {Kane},
  {Kains}, {Marquette}, {Martin}, {Martioli}, {Meintjes}, {Menzies},
  {Pedretti}, {Pollard}, {Sahu}, {Vinter}, {Wambsganss}, {Watson}, {Williams},
  {Zub}, {FUN Collaboration}, {Allen}, {Bolt}, {Bos}, {DePoy}, {Drummond},
  {Eastman}, {Gal-Yam}, {Gorbikov}, {Higgins}, {Janczak}, {Kaspi}, {Lee},
  {Mallia}, {Maury}, {Monard}, {Moorhouse}, {Morgan}, {Natusch}, {Ofek},
  {Park}, {Pogge}, {Polishook}, {Santallo}, {Shporer}, {Spector}, {Thornley},
  {Yee}, {MiNDSTEp Consortium}, {Bozza}, {Browne}, {Dominik}, {Dreizler},
  {Finet}, {Glitrup}, {Grundahl}, {Harps{\o}e}, {Hessman}, {Hinse},
  {Hundertmark}, {J{\o}rgensen}, {Liebig}, {Maier}, {Mancini}, {Mathiasen},
  {Rahvar}, {Ricci}, {Scarpetta}, {Southworth}, {Surdej}, {Zimmer}, {RoboNet
  Collaboration}, {Allan}, {Bramich}, {Snodgrass}, {Steele}, \&
  {Street}}]{Batista2011}
{Batista}, V., {Gould}, A., {Dieters}, S., {et~al.} 2011, \aap, 529, A102

\bibitem[{{Beaulieu} {et~al.}(2010){Beaulieu}, {Bennett}, {Batista}, {Cassan},
  {Kubas}, {Fouqu{\'e}}, {Kerrins}, {Mao}, {Miralda-Escud{\'e}}, {Wambsganss},
  {Gaudi}, {Gould}, \& {Dong}}]{Euclid2010}
{Beaulieu}, J.~P., {Bennett}, D.~P., {Batista}, V., {et~al.} 2010, in
  Astronomical Society of the Pacific Conference Series, Vol. 430, Pathways
  Towards Habitable Planets, ed. V.~{Coud{\'e} du Foresto}, D.~M. {Gelino}, \&
  I.~{Ribas}, 266

\bibitem[{{Bennett}(2010)}]{Bennett2010}
{Bennett}, D.~P. 2010, \apj, 716, 1408

\bibitem[{{Bonfils} {et~al.}(2013){Bonfils}, {Delfosse}, {Udry}, {Forveille},
  {Mayor}, {Perrier}, {Bouchy}, {Gillon}, {Lovis}, {Pepe}, {Queloz}, {Santos},
  {S{\'e}gransan}, \& {Bertaux}}]{Bonfils2013}
{Bonfils}, X., {Delfosse}, X., {Udry}, S., {et~al.} 2013, \aap, 549, A109

\bibitem[{{Bozza}(2010)}]{Bozza2010}
{Bozza}, V. 2010, \mnras, 408, 2188

\bibitem[{{Bozza} {et~al.}(2012){Bozza}, {Dominik}, {Rattenbury},
  {J{\o}rgensen}, {Tsapras}, {Bramich}, {Udalski}, {Bond}, {Liebig}, {Cassan},
  {Fouqu{\'e}}, {Fukui}, {Hundertmark}, {Shin}, {Lee}, {Choi}, {Park}, {Gould},
  {Allan}, {Mao}, {Wyrzykowski}, {Street}, {Buckley}, {Nagayama}, {Mathiasen},
  {Hinse}, {Novati}, {Harps{\o}e}, {Mancini}, {Scarpetta}, {Anguita},
  {Burgdorf}, {Horne}, {Hornstrup}, {Kains}, {Kerins}, {Kj{\ae}rgaard}, {Masi},
  {Rahvar}, {Ricci}, {Snodgrass}, {Southworth}, {Steele}, {Surdej},
  {Th{\"o}ne}, {Wambsganss}, {Zub}, {Albrow}, {Batista}, {Beaulieu}, {Bennett},
  {Caldwell}, {Cole}, {Cook}, {Coutures}, {Dieters}, {Prester}, {Donatowicz},
  {Greenhill}, {Kane}, {Kubas}, {Marquette}, {Martin}, {Menzies}, {Pollard},
  {Sahu}, {Williams}, {Szyma{\'n}ski}, {Kubiak}, {Pietrzy{\'n}ski},
  {Soszy{\'n}ski}, {Poleski}, {Ulaczyk}, {DePoy}, {Dong}, {Han}, {Janczak},
  {Lee}, {Pogge}, {Abe}, {Furusawa}, {Hearnshaw}, {Itow}, {Kilmartin},
  {Korpela}, {Lin}, {Ling}, {Masuda}, {Matsubara}, {Miyake}, {Muraki},
  {Ohnishi}, {Perrott}, {Saito}, {Skuljan}, {Sullivan}, {Sumi}, {Suzuki},
  {Sweatman}, {Tristram}, {Wada}, {Yock}, {Gulbis}, {Hashimoto}, {Kniazev}, \&
  {Vaisanen}}]{Bozza2012}
{Bozza}, V., {Dominik}, M., {Rattenbury}, N.~J., {et~al.} 2012, \mnras, 424,
  902

\bibitem[{{Bressan} {et~al.}(2012){Bressan}, {Marigo}, {Girardi}, {Salasnich},
  {Dal Cero}, {Rubele}, \& {Nanni}}]{Bressan2012}
{Bressan}, A., {Marigo}, P., {Girardi}, L., {et~al.} 2012, \mnras, 427, 127

\bibitem[{{Burrows} {et~al.}(2001){Burrows}, {Hubbard}, {Lunine}, \&
  {Liebert}}]{Burrows2001}
{Burrows}, A., {Hubbard}, W.~B., {Lunine}, J.~I., \& {Liebert}, J. 2001,
  Reviews of Modern Physics, 73, 719

\bibitem[{{Carroll} \& {Ostlie}(2006)}]{ModernAstro2006book}
{Carroll}, B.~W. \& {Ostlie}, D.~A. 2006, {An introduction to modern
  astrophysics and cosmology} (Addison-Wesley)

\bibitem[{{Cassan}(2008)}]{Cassan2008}
{Cassan}, A. 2008, \aap, 491, 587

\bibitem[{Cassan {et~al.}(2004)Cassan, Beaulieu, Brillant, Coutures, Dominik,
  Donatowicz, Jrgensen, Kubas, Albrow, Caldwell, P., Greenhill, Hill, Horne,
  Kane, Martin, Menzies, Pollard, Sahu, Vinter, Wambsganss, Watson, Williams,
  Fendt, Hauschildt, Heinmueller, Marquette, \& Thurl}]{Cassan69letter}
Cassan, A., Beaulieu, J.~P., Brillant, S., {et~al.} 2004, \aap, 419, L1

\bibitem[{{Cassan} {et~al.}(2006){Cassan}, {Beaulieu}, {Fouqu{\'e}},
  {Brillant}, {Dominik}, {Greenhill}, {Heyrovsk{\'y}}, {Horne}, {J{\o}rgensen},
  {Kubas}, {Stempels}, {Vinter}, {Albrow}, {Bennett}, {Caldwell}, {Calitz},
  {Cook}, {Coutures}, {Dominis}, {Donatowicz}, {Hill}, {Hoffman}, {Kane},
  {Marquette}, {Martin}, {Meintjes}, {Menzies}, {Miller}, {Pollard}, {Sahu},
  {Wambsganss}, {Williams}, {Udalski}, {Szyma{\'n}ski}, {Kubiak},
  {Pietrzy{\'n}ski}, {Soszy{\'n}ski}, {{\.Z}ebru{\'n}}, {Szewczyk}, \&
  {Wyrzykowski}}]{Cassan254}
{Cassan}, A., {Beaulieu}, J.-P., {Fouqu{\'e}}, P., {et~al.} 2006, \aap, 460,
  277

\bibitem[{{Cassan} {et~al.}(2010){Cassan}, {Horne}, {Kains}, {Tsapras}, \&
  {Browne}}]{Cassan2010}
{Cassan}, A., {Horne}, K., {Kains}, N., {Tsapras}, Y., \& {Browne}, P. 2010,
  \aap, 515, A52

\bibitem[{{Cassan} {et~al.}(2012){Cassan}, {Kubas}, {Beaulieu}, {Dominik},
  {Horne}, {Greenhill}, {Wambsganss}, {Menzies}, {Williams}, {J{\o}rgensen},
  {Udalski}, {Bennett}, {Albrow}, {Batista}, {Brillant}, {Caldwell}, {Cole},
  {Coutures}, {Cook}, {Dieters}, {Prester}, {Donatowicz}, {Fouqu{\'e}}, {Hill},
  {Kains}, {Kane}, {Marquette}, {Martin}, {Pollard}, {Sahu}, {Vinter},
  {Warren}, {Watson}, {Zub}, {Sumi}, {Szyma{\'n}ski}, {Kubiak}, {Poleski},
  {Soszynski}, {Ulaczyk}, {Pietrzy{\'n}ski}, \& {Wyrzykowski}}]{Cassan2012}
{Cassan}, A., {Kubas}, D., {Beaulieu}, J.-P., {et~al.} 2012, \nat, 481, 167

\bibitem[{{Choi} {et~al.}(2013){Choi}, {Han}, {Udalski}, {Sumi}, {Gaudi},
  {Gould}, {Bennett}, {Dominik}, {Beaulieu}, {Tsapras}, {Bozza}, {Abe}, {Bond},
  {Botzler}, {Chote}, {Freeman}, {Fukui}, {Furusawa}, {Itow}, {Ling}, {Masuda},
  {Matsubara}, {Miyake}, {Muraki}, {Ohnishi}, {Rattenbury}, {Saito},
  {Sullivan}, {Suzuki}, {Sweatman}, {Suzuki}, {Takino}, {Tristram}, {Wada},
  {Yock}, {The MOA Collaboration}, {Szyma{\'n}ski}, {Kubiak},
  {Pietrzy{\'n}ski}, {Soszy{\'n}ski}, {Skowron}, {Koz{\l}owski}, {Poleski},
  {Ulaczyk}, {Wyrzykowski}, {Pietrukowicz}, {The OGLE Collaboration},
  {Almeida}, {DePoy}, {Dong}, {Gorbikov}, {Jablonski}, {Henderson}, {Hwang},
  {Janczak}, {Jung}, {Kaspi}, {Lee}, {Malamud}, {Maoz}, {McGregor},
  {Mu{\~n}oz}, {Park}, {Park}, {Pogge}, {Shvartzvald}, {Shin}, {Yee}, {The
  {$\mu$}FUN Collaboration}, {Alsubai}, {Browne}, {Burgdorf}, {Calchi Novati},
  {Dodds}, {Fang}, {Finet}, {Glitrup}, {Grundahl}, {Gu}, {Hardis},
  {Harps{\o}e}, {Hinse}, {Hornstrup}, {Hundertmark}, {Jessen-Hansen},
  {Jrgensen}, {Kains}, {Kerins}, {Liebig}, {Lund}, {Lundkvist}, {Maier},
  {Mancini}, {Mathiasen}, {Penny}, {Rahvar}, {Ricci}, {Scarpetta}, {Skottfelt},
  {Snodgrass}, {Southworth}, {Surdej}, {Tregloan-Reed}, {Wambsganss}, {Wertz},
  {Zimmer}, {MiNDSTEp Consortium}, {Albrow}, {Bachelet}, {Batista}, {Brillant},
  {Cassan}, {Cole}, {Coutures}, {Dieters}, {Dominis Prester}, {Donatowicz},
  {Fouqu{\'e}}, {Greenhill}, {Kubas}, {Marquette}, {Menzies}, {Sahu}, {Zub},
  {The PLANET Collaboration}, {Bramich}, {Horne}, {Steele}, {Street}, \& {The
  RoboNet Collaboration}}]{Choi2013}
{Choi}, J.-Y., {Han}, C., {Udalski}, A., {et~al.} 2013, \apj, 768, 129

\bibitem[{{Claret} \& {Bloemen}(2011)}]{Claret2011}
{Claret}, A. \& {Bloemen}, S. 2011, \aap, 529, A75

\bibitem[{{Deleuil} {et~al.}(2008){Deleuil}, {Deeg}, {Alonso}, {Bouchy},
  {Rouan}, {Auvergne}, {Baglin}, {Aigrain}, {Almenara}, {Barbieri}, {Barge},
  {Bruntt}, {Bord{\'e}}, {Collier Cameron}, {Csizmadia}, {de La Reza},
  {Dvorak}, {Erikson}, {Fridlund}, {Gandolfi}, {Gillon}, {Guenther}, {Guillot},
  {Hatzes}, {H{\'e}brard}, {Jorda}, {Lammer}, {L{\'e}ger}, {Llebaria},
  {Loeillet}, {Mayor}, {Mazeh}, {Moutou}, {Ollivier}, {P{\"a}tzold}, {Pont},
  {Queloz}, {Rauer}, {Schneider}, {Shporer}, {Wuchterl}, \&
  {Zucker}}]{Deleuil2008}
{Deleuil}, M., {Deeg}, H.~J., {Alonso}, R., {et~al.} 2008, \aap, 491, 889

\bibitem[{{Devillard}(1999)}]{Devillard1999}
{Devillard}, N. 1999, in Astronomical Society of the Pacific Conference Series,
  Vol. 172, Astronomical Data Analysis Software and Systems VIII, ed. D.~M.
  {Mehringer}, R.~L. {Plante}, \& D.~A. {Roberts}, 333

\bibitem[{{D{\'{\i}}az} {et~al.}(2013){D{\'{\i}}az}, {Damiani}, {Deleuil},
  {Almenara}, {Moutou}, {Barros}, {Bonomo}, {Bouchy}, {Bruno}, {H{\'e}brard},
  {Montagnier}, \& {Santerne}}]{Diaz2013}
{D{\'{\i}}az}, R.~F., {Damiani}, C., {Deleuil}, M., {et~al.} 2013, \aap, 551,
  L9

\bibitem[{{Diolaiti} {et~al.}(2000){Diolaiti}, {Bendinelli}, {Bonaccini},
  {Close}, {Currie}, \& {Parmeggiani}}]{2000SPIE.4007..879D}
{Diolaiti}, E., {Bendinelli}, O., {Bonaccini}, D., {et~al.} 2000, in Presented
  at the Society of Photo-Optical Instrumentation Engineers (SPIE) Conference,
  Vol. 4007, Proc. SPIE Vol. 4007, p. 879-888, Adaptive Optical Systems
  Technology, Peter L. Wizinowich; Ed., ed. P.~L. {Wizinowich}, 879--888

\bibitem[{{Dominik}(2006)}]{Dominik2006}
{Dominik}, M. 2006, \mnras, 367, 669

\bibitem[{{Dominik}(2007)}]{Dominik2007cont}
{Dominik}, M. 2007, \mnras, 377, 1679

\bibitem[{{Dong} {et~al.}(2006){Dong}, {DePoy}, {Gaudi}, {Gould}, {Han},
  {Park}, {Pogge}, {Udalski}, {Szewczyk}, {Kubiak}, {Szyma{\'n}ski},
  {Pietrzy{\'n}ski}, {Soszy{\'n}ski}, {Wyrzykowski}, \&
  {{\.Z}ebru{\'n}}}]{Dong2006}
{Dong}, S., {DePoy}, D.~L., {Gaudi}, B.~S., {et~al.} 2006, \apj, 642, 842

\bibitem[{{Dong} {et~al.}(2009)}]{Dong2009}
{Dong}, S. {et~al.} 2009, \apj, 698, 1826

\bibitem[{{Duquennoy} \& {Mayor}(1991)}]{Duquennoy1991}
{Duquennoy}, A. \& {Mayor}, M. 1991, \aap, 248, 485

\bibitem[{{Geweke}(1992)}]{Geweke1992}
{Geweke}, J. 1992, in Bernardo, J.~M., Berger, J.~O., Dawid, A.~P. and Smith,
  A.~F.~M. (eds), Bayesian Statistics (Oxford University Press, New York),
  169--193

\bibitem[{{Gould}(2004)}]{Gould2004}
{Gould}, A. 2004, \apj, 606, 319

\bibitem[{{Gould}(2008)}]{Gould2008}
{Gould}, A. 2008, \apj, 681, 1593

\bibitem[{{Gould} {et~al.}(2010{\natexlab{a}}){Gould}, {Dong}, {Bennett},
  {Bond}, {Udalski}, \& {Kozlowski}}]{Gould2010}
{Gould}, A., {Dong}, S., {Bennett}, D.~P., {et~al.} 2010{\natexlab{a}}, \apj,
  710, 1800

\bibitem[{{Gould} {et~al.}(2010{\natexlab{b}}){Gould}, {Dong}, {Gaudi},
  {Udalski}, {Bond}, {Greenhill}, {Street}, {Dominik}, {Sumi}, {Szyma{\'n}ski},
  {Han}, {Allen}, {Bolt}, {Bos}, {Christie}, {DePoy}, {Drummond}, {Eastman},
  {Gal-Yam}, {Higgins}, {Janczak}, {Kaspi}, {Koz{\l}owski}, {Lee}, {Mallia},
  {Maury}, {Maoz}, {McCormick}, {Monard}, {Moorhouse}, {Morgan}, {Natusch},
  {Ofek}, {Park}, {Pogge}, {Polishook}, {Santallo}, {Shporer}, {Spector},
  {Thornley}, {Yee}, {{$\mu$}FUN Collaboration}, {Kubiak}, {Pietrzy{\'n}ski},
  {Soszy{\'n}ski}, {Szewczyk}, {Wyrzykowski}, {Ulaczyk}, {Poleski}, {OGLE
  Collaboration}, {Abe}, {Bennett}, {Botzler}, {Douchin}, {Freeman}, {Fukui},
  {Furusawa}, {Hearnshaw}, {Hosaka}, {Itow}, {Kamiya}, {Kilmartin}, {Korpela},
  {Lin}, {Ling}, {Makita}, {Masuda}, {Matsubara}, {Miyake}, {Muraki}, {Nagaya},
  {Nishimoto}, {Ohnishi}, {Okumura}, {Perrott}, {Philpott}, {Rattenbury},
  {Saito}, {Sako}, {Sullivan}, {Sweatman}, {Tristram}, {von Seggern}, {Yock},
  {MOA Collaboration}, {Albrow}, {Batista}, {Beaulieu}, {Brillant}, {Caldwell},
  {Calitz}, {Cassan}, {Cole}, {Cook}, {Coutures}, {Dieters}, {Dominis Prester},
  {Donatowicz}, {Fouqu{\'e}}, {Hill}, {Hoffman}, {Jablonski}, {Kane}, {Kains},
  {Kubas}, {Marquette}, {Martin}, {Martioli}, {Meintjes}, {Menzies},
  {Pedretti}, {Pollard}, {Sahu}, {Vinter}, {Wambsganss}, {Watson}, {Williams},
  {Zub}, {PLANET Collaboration}, {Allan}, {Bode}, {Bramich}, {Burgdorf},
  {Clay}, {Fraser}, {Hawkins}, {Horne}, {Kerins}, {Lister}, {Mottram},
  {Saunders}, {Snodgrass}, {Steele}, {Tsapras}, {RoboNet Collaboration},
  {J{\o}rgensen}, {Anguita}, {Bozza}, {Calchi Novati}, {Harps{\o}e}, {Hinse},
  {Hundertmark}, {Kj{\ae}rgaard}, {Liebig}, {Mancini}, {Masi}, {Mathiasen},
  {Rahvar}, {Ricci}, {Scarpetta}, {Southworth}, {Surdej}, {Th{\"o}ne}, \&
  {MiNDSTEp Consortium}}]{Gould4years2010}
{Gould}, A., {Dong}, S., {Gaudi}, B.~S., {et~al.} 2010{\natexlab{b}}, \apj,
  720, 1073

\bibitem[{{Gould} \& {Gaucherel}(1997)}]{GouldGaucherel1997}
{Gould}, A. \& {Gaucherel}, C. 1997, \apj, 477, 580

\bibitem[{{Gould} \& {Loeb}(1992)}]{Gould1992}
{Gould}, A. \& {Loeb}, A. 1992, \apj, 396, 104

\bibitem[{{Gould} {et~al.}(2009){Gould}, {Udalski}, {Monard}, {Horne}, {Dong},
  {Miyake}, {Sahu}, {Bennett}, {Wyrzykowski}, {Soszy{\'n}ski}, {Szyma{\'n}ski},
  {Kubiak}, {Pietrzy{\'n}ski}, {Szewczyk}, {Ulaczyk}, {OGLE Collaboration},
  {Allen}, {Christie}, {DePoy}, {Gaudi}, {Han}, {Lee}, {McCormick}, {Natusch},
  {Park}, {Pogge}, {{$\mu$}FUN Collaboration}, {Allan}, {Bode}, {Bramich},
  {Burgdorf}, {Dominik}, {Fraser}, {Kerins}, {Mottram}, {Snodgrass}, {Steele},
  {Street}, {Tsapras}, {RoboNet Collaboration}, {Abe}, {Bond}, {Botzler},
  {Fukui}, {Furusawa}, {Hearnshaw}, {Itow}, {Kamiya}, {Kilmartin}, {Korpela},
  {Lin}, {Ling}, {Masuda}, {Matsubara}, {Muraki}, {Nagaya}, {Ohnishi},
  {Okumura}, {Perrott}, {Rattenbury}, {Saito}, {Sako}, {Skuljan}, {Sullivan},
  {Sumi}, {Sweatman}, {Tristram}, {Yock}, {MOA Collaboration}, {Albrow},
  {Beaulieu}, {Coutures}, {Calitz}, {Caldwell}, {Fouque}, {Martin}, {Williams},
  \& {PLANET Collaboration}}]{Gould2009}
{Gould}, A., {Udalski}, A., {Monard}, B., {et~al.} 2009, \apjl, 698, L147

\bibitem[{{Guillot} {et~al.}(2012){Guillot}, {Lin}, \& {Morel}}]{Guillot2012}
{Guillot}, T., {Lin}, D.~N.~C., \& {Morel}, P. 2012, in American Astronomical
  Society Meeting Abstracts, Vol. 220, American Astronomical Society Meeting
  Abstracts \#220, \#121.03

\bibitem[{{Han} {et~al.}(2013){Han}, {Jung}, {Udalski}, {Sumi}, {Gaudi},
  {Gould}, {Bennett}, {Tsapras}, {Szyma{\'n}ski}, {Kubiak}, {Pietrzy{\'n}ski},
  {Soszy{\'n}ski}, {Skowron}, {Koz{\l}owski}, {Poleski}, {Ulaczyk},
  {Wyrzykowski}, {Pietrukowicz}, {The OGLE Collaboration}, {Abe}, {Bond},
  {Botzler}, {Chote}, {Freeman}, {Fukui}, {Furusawa}, {Harris}, {Itow}, {Ling},
  {Masuda}, {Matsubara}, {Muraki}, {Ohnishi}, {Rattenbury}, {Saito},
  {Sullivan}, {Sweatman}, {Suzuki}, {Tristram}, {Wada}, {Yock}, {The MOA
  Collaboration}, {Batista}, {Christie}, {Choi}, {DePoy}, {Dong}, {Hwang},
  {Kavka}, {Lee}, {Monard}, {Natusch}, {Ngan}, {Park}, {Pogge}, {Porritt},
  {Shin}, {Tan}, {Yee}, {The {$\mu$}FUN Collaboration}, {Alsubai}, {Bozza},
  {Bramich}, {Browne}, {Dominik}, {Horne}, {Hundertmark}, {Ipatov}, {Kains},
  {Liebig}, {Snodgrass}, {Steele}, {Street}, \& {The RoboNet
  Collaboration}}]{Han2013}
{Han}, C., {Jung}, Y.~K., {Udalski}, A., {et~al.} 2013, \apj, 778, 38

\bibitem[{{Janczak} {et~al.}(2010){Janczak}, {Fukui}, {Dong}, {Monard},
  {Koz{\l}owski}, {Gould}, {Beaulieu}, {Kubas}, {Marquette}, {Sumi}, {Bond},
  {Bennett}, {Abe}, {Furusawa}, {Hearnshaw}, {Hosaka}, {Itow}, {Kamiya},
  {Korpela}, {Kilmartin}, {Lin}, {Ling}, {Makita}, {Masuda}, {Matsubara},
  {Miyake}, {Muraki}, {Nagaya}, {Nagayama}, {Nishimoto}, {Ohnishi}, {Perrott},
  {Rattenbury}, {Sako}, {Saito}, {Skuljan}, {Sullivan}, {Sweatman}, {Tristram},
  {Yock}, {MOA Collaboration}, {An}, {Christie}, {Chung}, {DePoy}, {Gaudi},
  {Han}, {Lee}, {Mallia}, {Natusch}, {Park}, {Pogge}, {{$\mu$}FUN
  Collaboration}, {Anguita}, {Calchi Novati}, {Dominik}, {J{\o}rgensen},
  {Masi}, {Mathiasen}, {MiNDSTEp Collaboration}, {Batista}, {Brillant},
  {Cassan}, {Cole}, {Corrales}, {Coutures}, {Dieters}, {Fouqu{\'e}},
  {Greenhill}, \& {PLANET Collaboration}}]{Janczak2010}
{Janczak}, J., {Fukui}, A., {Dong}, S., {et~al.} 2010, \apj, 711, 731

\bibitem[{{Johnson} {et~al.}(2011){Johnson}, {Apps}, {Gazak}, {Crepp},
  {Crossfield}, {Howard}, {Marcy}, {Morton}, {Chubak}, \&
  {Isaacson}}]{Johnson2011}
{Johnson}, J.~A., {Apps}, K., {Gazak}, J.~Z., {et~al.} 2011, \apj, 730, 79

\bibitem[{{Jung} {et~al.}(2015){Jung}, {Udalski}, {Sumi}, {Han}, {Gould},
  {Skowron}, {Koz{\l}owski}, {Poleski}, {Wyrzykowski}, {Szyma{\'n}ski},
  {Pietrzy{\'n}ski}, {Soszy{\'n}ski}, {Ulaczyk}, {Pietrukowicz}, {Mr{\'o}z},
  {Kubiak}, {OGLE Collaboration}, {Abe}, {Bennett}, {Bond}, {Botzler},
  {Freeman}, {Fukui}, {Fukunaga}, {Itow}, {Koshimoto}, {Larsen}, {Ling},
  {Masuda}, {Matsubara}, {Muraki}, {Namba}, {Ohnishi}, {Philpott},
  {Rattenbury}, {Saito}, {Sullivan}, {Suzuki}, {Tristram}, {Tsurumi}, {Wada},
  {Yamai}, {Yock}, {Yonehara}, {The MOA Collaboration}, {Albrow}, {Choi},
  {DePoy}, {Gaudi}, {Hwang}, {Lee}, {Park}, {Owen}, {Pogge}, {Shin}, {Yee}, \&
  {The {$\mu$}FUN Collaboration}}]{Jung2014}
{Jung}, Y.~K., {Udalski}, A., {Sumi}, T., {et~al.} 2015, \apj, 798, 123

\bibitem[{{Kains} {et~al.}(2012){Kains}, {Browne}, {Horne}, {Hundertmark}, \&
  {Cassan}}]{Kains2012}
{Kains}, N., {Browne}, P., {Horne}, K., {Hundertmark}, M., \& {Cassan}, A.
  2012, \mnras, 426, 2228

\bibitem[{{Kains} {et~al.}(2009){Kains}, {Cassan}, {Horne}, {Albrow},
  {Dieters}, {Fouqu{\'e}}, {Greenhill}, {Udalski}, {Zub}, {Bennett}, {Dominik},
  {Donatowicz}, {Kubas}, {Tsapras}, {Anguita}, {Batista}, {Beaulieu},
  {Brillant}, {Bode}, {Bramich}, {Burgdorf}, {Caldwell}, {Cook}, {Coutures},
  {Dominis Prester}, {J{\o}rgensen}, {Kane}, {Marquette}, {Martin}, {Menzies},
  {Pollard}, {Rattenbury}, {Sahu}, {Snodgrass}, {Steele}, {Vinter},
  {Wambsganss}, {Williams}, {Kubiak}, {Pietrzy{\'n}ski}, {Soszy{\'n}ski},
  {Szewczyk}, {Szyma{\'n}ski}, {Ulaczyk}, \& {Wyrzykowski}}]{Kains2009}
{Kains}, N., {Cassan}, A., {Horne}, K., {et~al.} 2009, \mnras, 395, 787

\bibitem[{{Kato} {et~al.}(2007){Kato}, {Nagashima}, {Nagayama}, {Kurita},
  {Koerwer}, {Kawai}, {Yamamuro}, {Zenno}, {Nishiyama}, {Baba}, {Kadowaki},
  {Haba}, {Hatano}, {Shimizu}, {Nishimura}, {Nagata}, {Sato}, {Murai},
  {Kawazu}, {Nakajima}, {Nakaya}, {Kandori}, {Kusakabe}, {Ishihara},
  {Kaneyasu}, {Hashimoto}, {Tamura}, {Tanab{\'e}}, {Ita}, {Matsunaga},
  {Nakada}, {Sugitani}, {Wakamatsu}, {Glass}, {Feast}, {Menzies}, {Whitelock},
  {Fourie}, {Stoffels}, {Evans}, \& {Hasegawa}}]{Kato2007}
{Kato}, D., {Nagashima}, C., {Nagayama}, T., {et~al.} 2007, \pasj, 59, 615

\bibitem[{{Kervella} \& {Fouqu{\'e}}(2008)}]{Kervella2008}
{Kervella}, P. \& {Fouqu{\'e}}, P. 2008, \aap, 491, 855

\bibitem[{{Kubas} {et~al.}(2012){Kubas}, {Beaulieu}, {Bennett}, {Cassan},
  {Cole}, {Lunine}, {Marquette}, {Dong}, {Gould}, {Sumi}, {Batista},
  {Fouqu{\'e}}, {Brillant}, {Dieters}, {Coutures}, {Greenhill}, {Bond},
  {Nagayama}, {Udalski}, {Pompei}, {N{\"u}rnberger}, \& {Le
  Bouquin}}]{Kubas2012}
{Kubas}, D., {Beaulieu}, J.~P., {Bennett}, D.~P., {et~al.} 2012, \aap, 540, A78

\bibitem[{{Kubas} {et~al.}(2005){Kubas}, {Cassan}, {Beaulieu}, {Coutures},
  {Dominik}, {Albrow}, {Brillant}, {Caldwell}, {Dominis}, {Donatowicz},
  {Fendt}, {Fouqu{\'e}}, {J{\o}rgensen}, {Greenhill}, {Hill}, {Heinm{\"u}ller},
  {Horne}, {Kane}, {Marquette}, {Martin}, {Menzies}, {Pollard}, {Sahu},
  {Vinter}, {Wambsganss}, {Watson}, {Williams}, \& {Thurl}}]{Kubas2005}
{Kubas}, D., {Cassan}, A., {Beaulieu}, J.~P., {et~al.} 2005, \aap, 435, 941

\bibitem[{{Lafreni{\`e}re} {et~al.}(2007){Lafreni{\`e}re}, {Doyon}, {Marois},
  {Nadeau}, {Oppenheimer}, {Roche}, {Rigaut}, {Graham}, {Jayawardhana},
  {Johnstone}, {Kalas}, {Macintosh}, \& {Racine}}]{Lafreniere2007}
{Lafreni{\`e}re}, D., {Doyon}, R., {Marois}, C., {et~al.} 2007, \apj, 670, 1367

\bibitem[{{Luhman} {et~al.}(2007){Luhman}, {Joergens}, {Lada}, {Muzerolle},
  {Pascucci}, \& {White}}]{Luhman2007}
{Luhman}, K.~L., {Joergens}, V., {Lada}, C., {et~al.} 2007, Protostars and
  Planets V, 443

\bibitem[{{Ma} \& {Ge}(2014)}]{Ma2014}
{Ma}, B. \& {Ge}, J. 2014, \mnras, 439, 2781

\bibitem[{{Mao} \& {Paczynski}(1991)}]{MaoPaczynski1991}
{Mao}, S. \& {Paczynski}, B. 1991, \apjl, 374, L37

\bibitem[{{Marcy} \& {Butler}(2000)}]{Marcy2000}
{Marcy}, G.~W. \& {Butler}, R.~P. 2000, \pasp, 112, 137

\bibitem[{{Mayor} {et~al.}(2011){Mayor}, {Marmier}, {Lovis}, {Udry},
  {S{\'e}gransan}, {Pepe}, {Benz}, {Bertaux}, {Bouchy}, {Dumusque}, {Lo Curto},
  {Mordasini}, {Queloz}, \& {Santos}}]{Mayor2011}
{Mayor}, M., {Marmier}, M., {Lovis}, C., {et~al.} 2011, ArXiv e-prints
  [\eprint[arXiv]{1109.2497}]

\bibitem[{{McCarthy} \& {Zuckerman}(2004)}]{McCarthy2004}
{McCarthy}, C. \& {Zuckerman}, B. 2004, \aj, 127, 2871

\bibitem[{{Metchev} \& {Hillenbrand}(2009)}]{Metchev2009}
{Metchev}, S.~A. \& {Hillenbrand}, L.~A. 2009, \apjs, 181, 62

\bibitem[{{Molli{\`e}re} \& {Mordasini}(2012)}]{MolMordasini2012}
{Molli{\`e}re}, P. \& {Mordasini}, C. 2012, \aap, 547, A105

\bibitem[{{Moutou} {et~al.}(2013){Moutou}, {Bonomo}, {Bruno}, {Montagnier},
  {Bouchy}, {Almenara}, {Barros}, {Deleuil}, {D{\'{\i}}az}, {H{\'e}brard}, \&
  {Santerne}}]{Moutou2013}
{Moutou}, C., {Bonomo}, A.~S., {Bruno}, G., {et~al.} 2013, \aap, 558, L6

\bibitem[{{Nataf} {et~al.}(2013){Nataf}, {Gould}, {Fouqu{\'e}}, {Gonzalez},
  {Johnson}, {Skowron}, {Udalski}, {Szyma{\'n}ski}, {Kubiak},
  {Pietrzy{\'n}ski}, {Soszy{\'n}ski}, {Ulaczyk}, {Wyrzykowski}, \&
  {Poleski}}]{Nataf2013}
{Nataf}, D.~M., {Gould}, A., {Fouqu{\'e}}, P., {et~al.} 2013, \apj, 769, 88

\bibitem[{{Park} {et~al.}(2015){Park}, {Udalski}, {Han}, {Poleski}, {Skowron},
  {Koz{\l}owski}, {Wyrzykowski}, {Szyma{\'n}ski}, {Pietrukowicz},
  {Pietrzy{\'n}ski}, {Soszy{\'n}ski}, \& {Ulaczyk}}]{Park2015arXiv}
{Park}, H., {Udalski}, A., {Han}, C., {et~al.} 2015, ArXiv e-prints
  [\eprint[arXiv]{1503.03197}]

\bibitem[{{Penny} {et~al.}(2013){Penny}, {Kerins}, {Rattenbury}, {Beaulieu},
  {Robin}, {Mao}, {Batista}, {Calchi Novati}, {Cassan}, {Fouqu{\'e}},
  {McDonald}, {Marquette}, {Tisserand}, \& {Zapatero Osorio}}]{PennyEuclid}
{Penny}, M.~T., {Kerins}, E., {Rattenbury}, N., {et~al.} 2013, \mnras, 434, 2

\bibitem[{{Persson} {et~al.}(1998){Persson}, {Murphy}, {Krzeminski}, {Roth}, \&
  {Rieke}}]{Persson1998}
{Persson}, S.~E., {Murphy}, D.~C., {Krzeminski}, W., {Roth}, M., \& {Rieke},
  M.~J. 1998, \aj, 116, 2475

\bibitem[{{Sahlmann} {et~al.}(2011){Sahlmann}, {S{\'e}gransan}, {Queloz},
  {Udry}, {Santos}, {Marmier}, {Mayor}, {Naef}, {Pepe}, \&
  {Zucker}}]{Sahlmann2011}
{Sahlmann}, J., {S{\'e}gransan}, D., {Queloz}, D., {et~al.} 2011, \aap, 525,
  A95

\bibitem[{{Schechter} {et~al.}(1993){Schechter}, {Mateo}, \& {Saha}}]{DoPhot}
{Schechter}, P.~L., {Mateo}, M., \& {Saha}, A. 1993, \pasp, 105, 1342

\bibitem[{{Scott}(1992)}]{Scott1992}
{Scott}, D.~W. 1992, {Multivariate Density Estimation: Theory, Practice, and
  Visualization} (John Wiley \& Sons, New York, Chicester)

\bibitem[{{Shin} {et~al.}(2012{\natexlab{a}}){Shin}, {Han}, {Choi}, {Udalski},
  {Sumi}, {Gould}, {Bozza}, {Dominik}, {Fouqu{\'e}}, {Horne}, {Szyma{\'n}ski},
  {Kubiak}, {Soszy{\'n}ski}, {Pietrzy{\'n}ski}, {Poleski}, {Ulaczyk},
  {Pietrukowicz}, {Koz{\l}owski}, {Skowron}, {Wyrzykowski}, {OGLE
  Collaboration}, {Abe}, {Bennett}, {Bond}, {Botzler}, {Chote}, {Freeman},
  {Fukui}, {Furusawa}, {Itow}, {Kobara}, {Ling}, {Masuda}, {Matsubara},
  {Miyake}, {Muraki}, {Ohmori}, {Ohnishi}, {Rattenbury}, {Saito}, {Sullivan},
  {Suzuki}, {Suzuki}, {Sweatman}, {Takino}, {Tristram}, {Wada}, {Yock}, {MOA
  Collaboration}, {Bramich}, {Snodgrass}, {Steele}, {Street}, {Tsapras},
  {RoboNet Collaboration}, {Alsubai}, {Browne}, {Burgdorf}, {Calchi Novati},
  {Dodds}, {Dreizler}, {Fang}, {Grundahl}, {Gu}, {Hardis}, {Harps{\o}e},
  {Hinse}, {Hornstrup}, {Hundertmark}, {Jessen-Hansen}, {J{\o}rgensen},
  {Kains}, {Kerins}, {Liebig}, {Lund}, {Lunkkvist}, {Mancini}, {Mathiasen},
  {Penny}, {Rahvar}, {Ricci}, {Scarpetta}, {Skottfelt}, {Southworth}, {Surdej},
  {Tregloan-Reed}, {Wambsganss}, {Wertz}, {MiNDSTEp Consortium}, {Almeida},
  {Batista}, {Christie}, {DePoy}, {Dong}, {Gaudi}, {Henderson}, {Jablonski},
  {Lee}, {McCormick}, {McGregor}, {Moorhouse}, {Natusch}, {Ngan}, {Park},
  {Pogge}, {Tan}, {Thornley}, {Yee}, {{$\mu$}FUN Collaboration}, {Albrow},
  {Bachelet}, {Beaulieu}, {Brillant}, {Cassan}, {Cole}, {Corrales}, {Coutures},
  {Dieters}, {Dominis Prester}, {Donatowicz}, {Greenhill}, {Kubas},
  {Marquette}, {Menzies}, {Sahu}, {Zub}, \& {PLANET Collaboration}}]{Shin2012a}
{Shin}, I.-G., {Han}, C., {Choi}, J.-Y., {et~al.} 2012{\natexlab{a}}, \apj,
  755, 91

\bibitem[{{Shin} {et~al.}(2012{\natexlab{b}}){Shin}, {Han}, {Gould}, {Udalski},
  {Sumi}, {Dominik}, {Beaulieu}, {Tsapras}, {Bozza}, {Szyma{\'n}ski}, {Kubiak},
  {Soszy{\'n}ski}, {Pietrzy{\'n}ski}, {Poleski}, {Ulaczyk}, {Pietrukowicz},
  {Koz{\l}owski}, {Skowron}, {Wyrzykowski}, {OGLE Collaboration}, {Abe},
  {Bennett}, {Bond}, {Botzler}, {Freeman}, {Fukui}, {Furusawa}, {Hayashi},
  {Hearnshaw}, {Hosaka}, {Itow}, {Kamiya}, {Kilmartin}, {Kobara}, {Korpela},
  {Lin}, {Ling}, {Makita}, {Masuda}, {Matsubara}, {Miyake}, {Muraki}, {Nagaya},
  {Nishimoto}, {Ohnishi}, {Okumura}, {Omori}, {Perrott}, {Rattenbury}, {Saito},
  {Skuljan}, {Sullivan}, {Suzuki}, {Sweatman}, {Tristram}, {Wada}, {Yock}, {MOA
  Collaboration}, {Christie}, {Depoy}, {Dong}, {Gal-Yam}, {Gaudi}, {Hung},
  {Janczak}, {Kaspi}, {Maoz}, {McCormick}, {McGregor}, {Moorhouse},
  {Mu{\~n}oz}, {Natusch}, {Nelson}, {Pogge}, {Tan}, {Polishook}, {Shvartzvald},
  {Shporer}, {Thornley}, {Malamud}, {Yee}, {Choi}, {Jung}, {Park}, {Lee},
  {Park}, {Koo}, {{$\mu$}FUN Collaboration}, {Bajek}, {Bramich}, {Browne},
  {Horne}, {Ipatov}, {Snodgrass}, {Steele}, {Street}, {Alsubai}, {Burgdorf},
  {Calchi Novati}, {Dodds}, {Dreizler}, {Fang}, {Grundahl}, {Gu}, {Hardis},
  {Harps{\o}e}, {Hinse}, {Hundertmark}, {Jessen-Hansen}, {J{\o}rgensen},
  {Kains}, {Kerins}, {Liebig}, {Lund}, {Lundkvist}, {Mancini}, {Mathiasen},
  {Hornstrup}, {Penny}, {Proft}, {Rahvar}, {Ricci}, {Scarpetta}, {Skottfelt},
  {Southworth}, {Surdej}, {Tregloan-Reed}, {Wertz}, {Zimmer}, {Albrow},
  {Batista}, {Brillant}, {Caldwell}, {Calitz}, {Cassan}, {Cole}, {Cook},
  {Corrales}, {Coutures}, {Dieters}, {Dominis Prester}, {Donatowicz},
  {Fouqu{\'e}}, {Greenhill}, {Hill}, {Hoffman}, {Kane}, {Kubas}, {Marquette},
  {Martin}, {Meintjes}, {Menzies}, {Pollard}, {Sahu}, {Wambsganss}, {Williams},
  {Vinter}, \& {Zub}}]{Shin2012b}
{Shin}, I.-G., {Han}, C., {Gould}, A., {et~al.} 2012{\natexlab{b}}, \apj, 760,
  116

\bibitem[{{Silverman}(1986)}]{Silverman1986}
{Silverman}, B.~W. 1986, {Density Estimation for Statistics and Data Analysis}
  (Chapman and Hall, London)

\bibitem[{{Skowron} {et~al.}(2011){Skowron}, {Udalski}, {Gould}, {Dong},
  {Monard}, {Han}, {Nelson}, {McCormick}, {Moorhouse}, {Thornley}, {Maury},
  {Bramich}, {Greenhill}, {Koz{\l}owski}, {Bond}, {Poleski}, {Wyrzykowski},
  {Ulaczyk}, {Kubiak}, {Szyma{\'n}ski}, {Pietrzy{\'n}ski}, {Soszy{\'n}ski},
  {OGLE Collaboration}, {Gaudi}, {Yee}, {Hung}, {Pogge}, {DePoy}, {Lee},
  {Park}, {Allen}, {Mallia}, {Drummond}, {Bolt}, {{$\mu$}FUN Collaboration},
  {Allan}, {Browne}, {Clay}, {Dominik}, {Fraser}, {Horne}, {Kains}, {Mottram},
  {Snodgrass}, {Steele}, {Street}, {Tsapras}, {RoboNet Collaboration}, {Abe},
  {Bennett}, {Botzler}, {Douchin}, {Freeman}, {Fukui}, {Furusawa}, {Hayashi},
  {Hearnshaw}, {Hosaka}, {Itow}, {Kamiya}, {Kilmartin}, {Korpela}, {Lin},
  {Ling}, {Makita}, {Masuda}, {Matsubara}, {Muraki}, {Nagayama}, {Miyake},
  {Nishimoto}, {Ohnishi}, {Perrott}, {Rattenbury}, {Saito}, {Skuljan},
  {Sullivan}, {Sumi}, {Suzuki}, {Sweatman}, {Tristram}, {Wada}, {Yock}, {MOA
  Collaboration}, {Beaulieu}, {Fouqu{\'e}}, {Albrow}, {Batista}, {Brillant},
  {Caldwell}, {Cassan}, {Cole}, {Cook}, {Coutures}, {Dieters}, {Dominis
  Prester}, {Donatowicz}, {Kane}, {Kubas}, {Marquette}, {Martin}, {Menzies},
  {Sahu}, {Wambsganss}, {Williams}, {Zub}, \& {PLANET
  Collaboration}}]{Skowron2011}
{Skowron}, J., {Udalski}, A., {Gould}, A., {et~al.} 2011, \apj, 738, 87

\bibitem[{{Street} {et~al.}(2013){Street}, {Choi}, {Tsapras}, {Han},
  {Furusawa}, {Hundertmark}, {Gould}, {Sumi}, {Bond}, {Wouters}, {Zellem},
  {Udalski}, {The RoboNet Collaboration}, {Snodgrass}, {Horne}, {Dominik},
  {Browne}, {Kains}, {Bramich}, {Bajek}, {Steele}, {Ipatov}, {The MOA
  Collaboration}, {Abe}, {Bennett}, {Botzler}, {Chote}, {Freeman}, {Fukui},
  {Harris}, {Itow}, {Ling}, {Masuda}, {Matsubara}, {Miyake}, {Muraki},
  {Nagayama}, {Nishimaya}, {Ohnishi}, {Rattenbury}, {Saito}, {Sullivan},
  {Suzuki}, {Sweatman}, {Tristram}, {Wada}, {Yock}, {The OGLE Collaboration},
  {Szyma{\'n}ski}, {Kubiak}, {Pietrzy{\'n}ski}, {Soszy{\'n}ski}, {Poleski},
  {Ulaczyk}, {Wyrzykowski}, {The {$\mu$}FUN Collaboration}, {Yee}, {Dong},
  {Shin}, {Lee}, {Skowron}, {De Almeida}, {DePoy}, {Gaudi}, {Hung},
  {Jablonski}, {Kaspi}, {Klein}, {Hwang}, {Koo}, {Maoz}, {Mu{\~n}oz}, {Pogge},
  {Polishhook}, {Shporer}, {McCormick}, {Christie}, {Natusch}, {Allen},
  {Drummond}, {Moorhouse}, {Thornley}, {Knowler}, {Bos}, {Bolt}, {The PLANET
  Collaboration}, {Beaulieu}, {Albrow}, {Batista}, {Brillant}, {Caldwell},
  {Cassan}, {Cole}, {Corrales}, {Coutures}, {Dieters}, {Dominis Prester},
  {Donatowicz}, {Fouqu{\'e}}, {Bachelet}, {Greenhill}, {Kane}, {Kubas},
  {Marquette}, {Martin}, {Menzies}, {Pollard}, {Sahu}, {Wambsganss},
  {Williams}, {Zub}, {MiNDSTEp}, {Alsubai}, {Bozza}, {Burgdorf}, {Calchi
  Novati}, {Dodds}, {Dreizler}, {Finet}, {Gerner}, {Hardis}, {Harps{\o}e},
  {Hessman}, {Hinse}, {J{\o}rgensen}, {Kerins}, {Liebig}, {Mancini},
  {Mathiasen}, {Penny}, {Proft}, {Rahvar}, {Ricci}, {Scarpetta}, {Sch{\"a}fer},
  {Sch{\"o}nebeck}, {Southworth}, \& {Surdej}}]{Street2013}
{Street}, R.~A., {Choi}, J.-Y., {Tsapras}, Y., {et~al.} 2013, \apj, 763, 67

\bibitem[{{Sumi} {et~al.}(2011){Sumi}, {Kamiya}, {Bennett}, {Bond}, {Abe},
  {Botzler}, {Fukui}, {Furusawa}, {Hearnshaw}, {Itow}, {Kilmartin}, {Korpela},
  {Lin}, {Ling}, {Masuda}, {Matsubara}, {Miyake}, {Motomura}, {Muraki},
  {Nagaya}, {Nakamura}, {Ohnishi}, {Okumura}, {Perrott}, {Rattenbury}, {Saito},
  {Sako}, {Sullivan}, {Sweatman}, {Tristram}, {Udalski}, {Szyma{\'n}ski},
  {Kubiak}, {Pietrzy{\'n}ski}, {Poleski}, {Soszy{\'n}ski}, {Wyrzykowski},
  {Ulaczyk}, \& {Microlensing Observations in Astrophysics (MOA)
  Collaboration}}]{Sumi2011}
{Sumi}, T., {Kamiya}, K., {Bennett}, D.~P., {et~al.} 2011, \nat, 473, 349

\bibitem[{{Tsapras} {et~al.}(2009)}]{Tsapras2009}
{Tsapras}, Y. {et~al.} 2009, AN, 330, 4

\bibitem[{{Yee} {et~al.}(2014){Yee}, {Albrow}, {Barry}, {Bennett}, {Bryden},
  {Chung}, {Gaudi}, {Gehrels}, {Gould}, {Penny}, {Rattenbury}, {Ryu},
  {Skowron}, {Street}, \& {Sumi}}]{Yee2014Exopag}
{Yee}, J.~C., {Albrow}, M., {Barry}, R.~K., {et~al.} 2014, ArXiv e-prints
  [\eprint[arXiv]{1409.2759}]

\bibitem[{{Zub} {et~al.}(2011){Zub}, {Cassan}, {Heyrovsk{\'y}}, {Fouqu{\'e}},
  {Stempels}, {Albrow}, {Beaulieu}, {Brillant}, {Christie}, {Kains},
  {Koz{\l}owski}, {Kubas}, {Wambsganss}, {Batista}, {Bennett}, {Cook},
  {Coutures}, {Dieters}, {Dominik}, {Dominis Prester}, {Donatowicz},
  {Greenhill}, {Horne}, {J{\o}rgensen}, {Kane}, {Marquette}, {Martin},
  {Menzies}, {Pollard}, {Sahu}, {Vinter}, {Williams}, {Gould}, {Depoy},
  {Gal-Yam}, {Gaudi}, {Han}, {Lipkin}, {Maoz}, {Ofek}, {Park}, {Pogge},
  {McCormick}, {Udalski}, {Szyma{\'n}ski}, {Kubiak}, {Pietrzy{\'n}ski},
  {Soszy{\'n}ski}, {Szewczyk}, {Wyrzykowski}, \& {PLANET
  Collaboration}}]{Zub2011}
{Zub}, M., {Cassan}, A., {Heyrovsk{\'y}}, D., {et~al.} 2011, \aap, 525, A15

\end{thebibliography}

\clearpage

\end{document}